\documentclass[10pt,conference]{ieeetran}

\usepackage{booktabs}   
\usepackage{subcaption} 
\usepackage{amsmath,amsthm,amssymb}

\usepackage[table]{xcolor}

\usepackage{threeparttable}

\usepackage{wasysym}

\usepackage{listings}

\usepackage{tikz}

\usepackage{array,multirow}

\usepackage{stmaryrd}

\usepackage[noadjust]{cite}

\usepackage[multiple]{footmisc}

\usepackage[hyphens]{url}

\theoremstyle{definition}
\newtheorem{definition}{Definition}

\newif\ifdraft \drafttrue
\newif\iftext \textfalse
\newif\iflater \latertrue
\newif\ifspace \spacefalse
\newif\ifaftersubmission \aftersubmissionfalse
\newif\ifcameraready  \camerareadytrue
\newif\ifexceptions \exceptionsfalse

\makeatletter \@input{texdirectives} \makeatother

\usepackage{cite}
\usepackage{amsmath,amsfonts}
\usepackage{algorithmic}
\usepackage{hyperref}
\usepackage{graphicx}
\usepackage{textcomp}
\usepackage[capitalize]{cleveref}
\usepackage[inline]{enumitem}

\definecolor{darkgreen}{RGB}{0,150,0}

\usepackage[normalem]{ulem}

\usepackage{listings}

\usepackage{xspace}

\makeatletter
\begingroup
\lccode`\A=`\-
\lccode`\N=`\N
\lccode`\V=`\V
\lowercase{\endgroup\def\memory@noval{ANoValue-}}
\long\def\memory@fiBgb\fi#1#2{\fi}
\long\def\memory@fiTBb\fi#1#2#3{\fi#2}
\newcommand\memory@ifnovalF[1]
  {%
    \ifx\memory@noval#1%
      \memory@fiBgb
    \fi
    \@firstofone
  }
\newcommand\memory@ifnovalTF[1]
  {%
    \ifx\memory@noval#1%
      \memory@fiTBb
    \fi
    \@secondoftwo
  }
\newcommand\memory@Oarg[2]
  {%
    \@ifnextchar[{\memory@Oarg@{#2}}{#2{#1}}%
  }
\long\def\memory@Oarg@#1[#2]
  {%
    #1{#2}%
  }
\newcommand*\memory@oarg
  {%
    \memory@Oarg\memory@noval
  }
\newcommand*\memory@ifcoloropt
  {%
    \@ifnextchar[\memory@ifcoloropt@true\memory@ifcoloropt@false
  }
\long\def\memory@ifcoloropt@true#1\memory@noval#2#3
  {%
    #2%
  }
\long\def\memory@ifcoloropt@false#1\memory@noval#2#3
  {%
    #3%
  }
\newlength\memory@width
\newlength\memory@height
\newcount\memory@num
\newcommand*\memory@blocks[2]
  {%
    \memory@num#1\relax
    \fboxsep-\fboxrule
    \memory@ifcoloropt#2\memory@noval
      {\def\memory@color{\textcolor#2}}
      {\def\memory@color{\textcolor{#2}}}%
    \loop
    \ifnum\memory@num>0
      \fbox{\memory@color{\rule{\memory@width}{\memory@height}}}%
      \kern-\fboxrule
      \advance\memory@num\m@ne
    \repeat
  }
\newcommand*\memory
  {%
    \begingroup
    \memory@oarg\memory@a
  }
\newcommand*\memory@a[2]
  {%
    \memory@ifnovalF{#1}{\memory@width#1\relax}%
    \memory@Oarg\memory@height{\memory@b{#2}}%
  }
\newcommand*\memory@b[3]
  {%
    \memory@ifnovalF{#2}{\memory@height#2\relax}%
    \memory@oarg{\memory@c{#1}{#3}}%
  }
\newcommand*\memory@c[3]
  {%
    \memory@ifnovalTF{#3}
      {\ensuremath{\memory@blocks{#1}{#2}}}
      {\ensuremath{\underbrace{\memory@blocks{#1}{#2}}_{\text{#3}}}}%
    \endgroup
  }

\makeatother

\newcommand{\judgment}[3][]{
  {\centering
  \smallskip
  \begin{tabular}{c}
    #2 \\
    \hline
    #3
  \end{tabular}{\sc #1}
  \smallskip\par}}

\newcommand{\judgmentbr}[4][]{
  {\centering
  \smallskip
  \begin{tabular}{c}
    #2 \\
    #3 \\
    \hline
    #4
  \end{tabular}{\sc #1}
   \smallskip\par}}

\newcommand{\judgmenttwobrlong}[5][]{
  {
    \centering
    \smallskip
    \begin{tabular}{c c}
       #2 & #3 \\
       \multicolumn{2}{c}{#4} \\
       \hline
       \multicolumn{2}{c}{#5}
    \end{tabular}{\sc #1}
    \vspace{\belowdisplayskip}\par
  }}

\newcommand{\judgmenttwo}[4][]{
  {\centering
  \smallskip
  \begin{tabular}{c c}
    #2 & #3 \\
    \hline
    \multicolumn{2}{c}{#4}
  \end{tabular}{\sc #1}
  \smallskip\par}}

\newcommand{\judgmentthree}[5][]{
  {\centering
  \smallskip
  \begin{tabular}{c c c}
    #2 & #3 & #4 \\
    \hline
    \multicolumn{3}{c}{#5}
  \end{tabular}{\sc #1}
  \smallskip\par}}

\newcommand{\word}{w}
\newcommand{\addr}{a}
\newcommand{\WORDS}{{\mathcal W}}
\newcommand{\reg}{r}
\newcommand{\REGS}{{\mathcal R}}

\newcommand{\mach}{m}
\newcommand{\nach}{n}

\newcommand{\MACHS}{{\mathcal M}}

\newcommand{\obs}{e}
\newcommand{\obsT}{\mathcal{E}}

\newcommand{\OBSS}{\mathit{EVENTS}}

\newcommand{\PCname}{\textsc{pc}}
\newcommand{\SP}{\textsc{sp}}

\newcommand{\component}{k}
\newcommand{\components}{K}
\newcommand{\COMPONENTS}{{\mathcal K}}

\newcommand{\notfinished}[2]{#1 \cdot #2}

\newcommand{\ret}{\mathit{Ret}}
\newcommand{\intProp}{\mathit{Int}}
\newcommand{\confProp}{\mathit{Conf}}
\newcommand{\cconfProp}{\mathit{CConf}}
\newcommand{\cintProp}{\mathit{CInt}}

\newcommand{\stepstounder}[1]{\stackrel{\mbox{\tiny{$#1$}}}{\Longrightarrow}}

\newcommand{\emplist}{\varepsilon}

\newcommand{\unsealed}{\mathit{free}}
\newcommand{\sealed}{\mathit{sealed}}
\newcommand{\public}{\mathit{public}}
\newcommand{\object}{\mathit{active}}

\newcommand{\context}{\mathit{c}}
\newcommand{\CONTEXTS}{\mathit{C}}

\newcommand*{\tagNoDepth}{\textsc{unused}}
\newcommand*{\tagStackDepth}[1]{\textsc{stack} ~ #1}
\newcommand*{\tagPCDepth}[1]{\textsc{pc} ~ #1}

\newcommand{\labeledrow}[3]{{\tt #1} & {\tt #2} & #3\\}

\newcommand{\wbcf}{\textnormal{\sc WBCF}}
\newcommand{\clri}{\textnormal{\sc ClrI}}
\newcommand{\clrc}{\textnormal{\sc ClrC}}
\newcommand{\clei}{\textnormal{\sc CleI}}
\newcommand{\clec}{\textnormal{\sc CleC}}

\newcommand{\genbox}[3]{\tikz \filldraw[fill=#2] (0,0) rectangle (#1,#1) node[pos=.5]{#3};}
\newcommand{\freebox}{\genbox{10pt}{blue!30}{}}
\newcommand{\pubbox}{\genbox{10pt}{lightgray}{}}
\newcommand{\objbox}{\genbox{10pt}{yellow}{}}
\newcommand{\sealbox}{\genbox{10pt}{red}{}}

\newcommand{\leftvariant}{cyan!40}
\newcommand{\rightvariant}{orange!50}

\makeatletter
\newcommand{\linebreakand}{%
  \end{@IEEEauthorhalign}
  \hfill\mbox{}\par
  \mbox{}\hfill\begin{@IEEEauthorhalign}
}
\makeatother

\begin{document}

\title{Formalizing Stack Safety as a Security Property\ifcameraready\else\vspace{-0.75cm}\fi}

\ifcameraready
\author{
  \IEEEauthorblockN{
    Sean Noble Anderson
  }
  \IEEEauthorblockA{
    Portland State University\\
    ander28@pdx.edu\\
  }
  \and
  \IEEEauthorblockN{
    Roberto Blanco
  }
  \IEEEauthorblockA{
    MPI-SP\\
    roberto.blanco@mpi-sp.org\\
  }
  \and
  \IEEEauthorblockN{
    Leonidas Lampropoulos
  }
  \IEEEauthorblockA{
    University of Maryland, College Park\\
    leonidas@umd.edu\\
  }
  \linebreakand
  \IEEEauthorblockN{
    Benjamin C. Pierce
  }
  \IEEEauthorblockA{
    University of Pennsylvania\\
    bcpierce@cis.upenn.edu\\
   }
  \and
  \IEEEauthorblockN{
    Andrew Tolmach
  }
  \IEEEauthorblockA{
    Portland State University\\
    tolmach@pdx.edu\\
  }
}
\fi


\maketitle

\begin{abstract}

The term {\em stack safety} is used to describe a variety of compiler,
run-time, and hardware mechanisms for protecting stack memory. Unlike
``the heap,'' the ISA-level stack does not correspond to a single
high-level language concept: different compilers use it in different
ways to support procedural and functional abstraction mechanisms from
a wide range of languages.  This protean nature makes it difficult to
nail down what it means to correctly enforce stack safety.

We propose a new formal characterization of stack safety using concepts
from language-based security. Rather than treating
stack safety as a monolithic property, we decompose it into an
integrity property and a confidentiality property for each of the
caller and the callee, plus a control-flow property: five properties
in all.
This formulation is motivated by a particular class of enforcement
mechanisms, the ``lazy'' stack safety micro-policies studied by
Roessler and DeHon~\cite{DBLP:conf/sp/RoesslerD18}, which permit
functions to write into one another's frames but taint the
changed locations so that the frame's owner cannot access them. No
existing characterization of stack safety captures this style of
safety; we capture it here by stating our properties in terms of the
observable behavior of the system.

Our properties go further than previous formal definitions of stack
safety, supporting caller- and callee-saved registers, arguments
passed on the stack, \ifexceptions tail-call elimination, and
exceptions.  \else and tail-call elimination. \fi
%
We validate the properties by using them to distinguish between
correct and incorrect implementations of Roessler and DeHon's
micro-policies using property-based random testing. Our test harness
successfully identifies several broken variants, including Roessler
and DeHon's lazy policy; a repaired version of their policy passes
our tests.

\end{abstract}

\newcommand{\paragraphx}[1]{\emph{#1.}}

\section{Introduction}
\label{sec:intro}


Functions in high-level languages (and related abstractions such
as procedures, methods, etc.) are units of computation
that invoke one another to define larger computations in a modular way.
At a low level, each function activation manages its own local
variables, spilled temporaries, etc., as well as information about the
{caller} to which it will return.
The \emph{call stack} is the fundamental data structure used to
implement functions, aided by an Application Binary Interface (ABI)
that defines how registers are shared between activations.

From a security perspective, attacks on the call stack
are attacks on the function abstraction itself.
Indeed, the stack is an ancient~\cite{phrack96:smashingthestack} and
perennial~\cite{mitre-cwe,DBLP:conf/raid/VeendCB12,
  DBLP:conf/sp/SzekeresPWS13,
  DBLP:conf/sp/HuSACSL16,msrc-bluehat,chromium-security}
target for low-level attacks, sometimes involving control-flow
hijacking via corrupting the return address, sometimes memory corruption
more generally.
%
%

The variety in attacks on the stack is mirrored in the range of
software and hardware protections that aim to prevent them,
including stack canaries~\cite{Cowan+98},
bounds checking~\cite{NagarakatteZMZ09,NagarakatteZMZ10,DeviettiBMZ08},
split stacks~\cite{Kuznetsov+14},
shadow stacks~\cite{Dang+15,Shanbhogue+19},
capabilities~\cite{Woodruff+14,Chisnall+15,SkorstengaardLocal,SkorstengaardSTKJFP,Georges22:TempsDesCerises},
and hardware tagging~\cite{DBLP:conf/sp/RoesslerD18,Gollapudi+23}.
But enforcement mechanisms can be brittle, successfully eliminating one
attack while leaving room for others. To avoid an endless game of
whack-a-mole, we seek formal properties of safe behavior that can be
proven, or at least rigorously tested. Such properties can be used as
the specification against which enforcement can be validated---even
enforcement mechanisms that do {\em not} fulfill a property can benefit from the
ability to articulate why and when they may fail.

Many of the mechanisms listed above are fundamentally ill-suited for
offering formal guarantees: they may impede attackers, but they do not provide
universal protection. Shadow stacks, for instance, aim to ``restrict
the flexibility available in creating gadget chains''
\cite{Shanbhogue+19}, not to categorically rule out attacks. Other
mechanisms, such as SoftBound~\cite{NagarakatteZMZ09} and code-pointer
integrity~\cite{Kuznetsov+14}, do aim for stronger guarantees, but not
formal ones.  To our knowledge, the sole
line of work making a formal claim to protect stack safety is the
study of secure calling conventions by Skorstengaard et
al.~\cite{SkorstengaardSTKJFP} and Georges et
al.~\cite{Georges22:TempsDesCerises}.

Some of the other mechanisms listed above should also be amenable to strong
formal guarantees.  In particular, Roessler and DeHon
\cite{DBLP:conf/sp/RoesslerD18} present an array of tag-based
micro-policies~\cite{pump_oakland2015} for stack safety that aim to offer
universal protection.  But the reasoning involved can be subtle:
they include micro-policy optimizations, Lazy Tagging and Lazy Clearing
(likely to be deployed together, which we hereafter refer to as Lazy
Tagging and Clearing, or LTC). LTC allows function activations to write
improperly into one another's stack frames, but ensures that the owner of
the corrupted memory cannot access it afterward, avoiding expensive
clearings of the stack frame.
Under this policy, one function activation {\em can} corrupt another's
memory---just not in ways that affect observable behavior.
Therefore, LTC would not fulfill Georges et al.'s property
(adapted to the tagged setting).  But LTC does arguably enforce stack safety,
or as Roessler and DeHon describe it informally, a sort of data-flow
integrity tied to the stack. A looser, more observational
definition of stack safety is needed to fit this situation.

We propose here a formal characterization of stack safety based on
the intuition of protecting function activations from each other and
using the tools of language-based
security~\cite{sabelfeld2003language} to treat function activations as
security principals.  We decompose stack safety into a family of
properties describing
the {\em integrity} and {\em confidentiality} of the caller’s local state
and the callee's behavior during (and after) the callee's execution,
together with the
{\em well-bracketed control flow} (\(\wbcf\)) property articulated by
Skorstengaard et al.~\cite{SkorstengaardSTKJFP}.

Our properties are stated abstractly in the hope that they can also be
applied to other enforcement mechanisms besides LTC.
However, it does not seem feasible to give
a universal definition of stack safety
that applies to all architectures and compilers.
While many security properties can be described purely at
the level of a high-level programming language and translated to a
target machine by a secure compiler, stack safety cannot be defined in
this way, since ``the stack'' is not explicitly present in the
definitions of most source languages but rather is implicit in the semantics
of features such as calls and returns.\footnote{
Contrast Azevedo de
  Amorim et al.'s work on heap safety
  \cite{DBLP:conf/post/AmorimHP18}, where the concept of the heap figures
  directly in high-level language semantics and its security is
  therefore amenable to a high-level treatment.}
But neither can stack safety be described coherently as a purely
low-level property;
indeed, at the lowest level, the specification of a ``well-behaved
stack'' is almost vacuous. The ISA is not concerned with such
questions as whether a caller's frame should be readable or writable
to its callee. Those are the purview of high-level languages built
atop the hardware stack.

Thus, any low-level treatment of stack safety must begin by asking:
which high-level features are supported in a given
setting using the stack, and how does their presence influence
the expectation of well-bracketed control flow, confidentiality,
and integrity? We begin with a simple system
with very few features, then move to a more realistic one supporting
tail-call elimination, argument passing on the stack, and callee-save
registers. Our properties are factored so that the basic structure
of each of our five properties remains constant while the presence or
absence of different features leads to subtler differences in how
they behave.

We demonstrate the usefulness of our properties for distinguishing
between correct and
incorrect enforcement using
QuickChick~\cite{Denes:VSL2014,Pierce:SF4}, a property-based
random testing tool for Coq.
Indeed, we find that the published version of LTC is flawed in
a way that undermines both integrity and confidentiality; after
correcting this flaw, LTC satisfies all of our
properties.  Further, we modify LTC to protect the features of
our more realistic system and apply
random testing to validate this extended protection mechanism against
the extended properties.

In summary, we offer the following contributions:

\begin{itemize}
\item We give a novel characterization of stack safety as a conjunction
  of security properties---confidentiality and integrity for callee
  and caller---plus well-bracketed control-flow.
  The properties are parameterized over a notion of
  external observation, allowing them to characterize lazy enforcement
  mechanisms.
\item We extend these core definitions to
  describe a realistic setting with argument passing on the stack,
  callee-saves registers, and tail-call elimination. The model is
  modular enough that adding these features is straightforward.
\item We validate a published enforcement mechanism, \emph{Lazy
  Tagging and Clearing}, via property-based random testing, find that
it falls short, and propose and validate a fix.
\end{itemize}
The following section offers a brief overview of our framework and
assumptions. \cref{sec:example} walks through a function call in
a simple example machine, discusses informally how each of our properties
applies to it, and motivates the properties
from a security perspective.
\cref{sec:formal} formalizes the machine model,
its {\em security semantics}, and the stack safety properties built on these.
\Cref{sec:extensions} describes how to support an extended set of features.
\Cref{sec:enforcement} describes the micro-policies that we test,
\cref{sec:testing} the testing framework itself, and
\cref{sec:relwork,sec:future} related and future work.

The accompanying artifact \footnote{https://github.com/SNoAnd/stack-safety}
contains formal definitions (in Coq) of our
properties, plus our testing framework.  It does not include proofs, since
we use Coq primarily for the QuickChick testing library and to
ensure that our definitions are unambiguous.  Formal proofs are left
as future work.

\section{Framework and Assumptions}
\label{sec:ideas}

Stack safety properties need to describe the behavior of machine code, but they naturally
talk about function activations and stack contents---abstractions that
are typically not visible at machine level. To bridge this gap,
our properties are defined in terms of a {\em security semantics} layered on top of
the standard execution semantics of the machine.  The security semantics identifies certain
state transitions of the machine as {\em security-relevant operations}, which update
a notional {\em security context}.  This context consists of
an (abstract) stack of function activations, each associated with a {\em view}
that maps each machine {\em state element} (memory location or register)
to a {\em security class} (active, sealed, etc.) specifying how the activation
can access the element.
The action of a security-relevant operation on the context is defined by a function
that characterizes how the operation's underling machine code ought to
implement the function abstraction in terms of the stack and registers.

Given the security classes of the elements of the machine state, we
define high-level
security properties---integrity, confidentiality, and well-bracketed
control flow---as
predicates that must hold on each call. These predicates draw on the idea of
\emph{variant} states from the theory of non-interference, plus a notion of
{\em observable events}, which might include specific function calls (e.g., system
calls that perform I/O), writes to special addresses representing
memory-mapped regions, etc. For example, to show that certain locations are kept
secret, it suffices to compare executions starting at machine states which vary at those locations
and check that their traces of observable events are the same. This structure
allows us to talk about the eventual impact of leaks or memory corruption without
reference to internal implementation details and, in particular, to support lazy enforcement by
flagging corruption of values only when it can actually impact visible behavior.

We introduce these properties by example in \cref{sec:example} and
formally in \cref{sec:formal}. In the remainder
of this section we introduce the underlying semantic framework
in more detail.

\paragraph*{Machine Model}
We assume a conventional ISA (e.g., RISC-V, x86-64, etc.), with registers including a program counter
and stack pointer.
We make no particular assumptions about the provenance of the machine code; in particular,
we do not assume any particular compiler.
If the machine is enhanced with 
enforcement mechanisms such as hardware
tags~\cite{pump_hasp2014,Gollapudi+23} or
capabilities~\cite{Woodruff+14}, we
assume that the behavior of these mechanisms is incorporated into the basic
step semantics of the machine, with a notion of ``compatible'' states that
share security behavior that may be defined based on the enforcement mechanism.
Failstop behavior by enforcement mechanisms is modeled as stepping to the same state
(and thus silently diverging).

\paragraph*{Security Semantics}
A security semantics extends the core machine model
with additional context about the identities of current and pending
functions (which act as security principals) and about their security
requirements on registers and memory. This added context is purely notional;
it does not affect the behavior of the core machine. The security context
evolves dynamically through the execution of security-relevant operations,
which include calls, returns, and frame manipulation.
Our security properties are phrased in terms of this context, often as predicates
on future states (``when control returns to the current function, X
must hold...'')
or as relations on traces of future execution
(hyper-properties).

Security-relevant operations abstract over the implementation details of the
actions they take. Since the same machine instruction may be used by compilers for
different purposes, we assume that the compiler or another trusted
source has provided
labels to identify the security-relevant purpose of each instruction,
if any. For instance,
in the tagged RISC-V architecture
that we use in our examples and tests,
calls and returns are conventionally performed using the {\tt jal}
(``jump-and-link'')
and {\tt jalr} (``jump-and-link-register'') instructions, but these
instructions might also be used for other things.

These considerations lead to an annotated version of the
machine transition function, written \(\mach \xrightarrow{\bar{\psi},
  \obs} \mach'\), where \(\mach\) and \(\mach\) are machine states, \(\obs\) is
an optional externally observable event, and \(\overline{\psi}\) is a
list of security-relevant operations---necessary because a single step might
perform multiple simultaneous operations.
This is then lifted into a transition between pairs of machine states
and contexts by applying a transition function parameterized by the operation.
We will decompose this function into rules associated with each operation and introduce
them as needed.
The most important of these rules describe call and return operations.
A call pushes a new view onto the context stack and changes the class of the
caller's data to protect it from the new callee; a return reverses these steps.
Other operations signal how parts of the stack frame are being used to store
or share data, and their corresponding rules alter the classes of different
state elements accordingly.

Exactly which operations and rules are needed depends on
what code features we wish to support.
The set of security-relevant operations (\(\Psi\)) covered in this paper is given in
\cref{tab:psi}. A core set of operations covering calls, returns, and local
memory is introduced in the example in \cref{sec:example}
and formalized in \cref{sec:formal}. An extended set covering simple memory sharing and
tail-call elimination is described in \cref{sec:extensions} and tested in \cref{sec:testing}.
The remaining operations are needed for the capability-based model in
\cref{app:ptr}.

\newcommand{\example}{\rowcolor{black!0}}
\newcommand{\testing}{\rowcolor{black!10}}
\newcommand{\theory}{\rowcolor{black!25}}

\begin{table}
\begin{center}
  \begin{tabular}{| l | l | l |}
    \hline
    Operation \(\psi \in \Psi\) & Parameters & Sections\\
    \hline
    \example \(\mathbf{call}\) & target address, argument registers & \ref{sec:example},\ref{sec:formal}\\
    \testing & stack arguments (base, offset \& size) & \ref{sec:extensions},\ref{sec:testing} \\
    \example \(\mathbf{return}\) & & \ref{sec:example},\ref{sec:formal}\\
    \example \(\mathbf{alloc}\) & offset \& size & \ref{sec:example},\ref{sec:formal}\\
    \testing & public flag & \ref{sec:extensions},\ref{sec:testing} \\
    \example \(\mathbf{dealloc}\) & offset \& size &  \ref{sec:example},\ref{sec:formal}\\
    \testing \(\mathbf{tail call}\) & (same as for \(\mathbf{call}\)) & \ref{sec:extensions},\ref{sec:testing} \\
    \theory \(\mathbf{promote}\) & register, offset \& size & \ref{app:ptr} \\
    \theory \(\mathbf{propagate}\) & source register/address & \ref{app:ptr} \\
    \theory & destination register/address & \ref{app:ptr} \\
    \theory \(\mathbf{clear}\) & target register/address & \ref{app:ptr} \\
    \hline
  \end{tabular}
\end{center}
\caption{Security-relevant operations and their parameters, with the
  sections where they are first defined or used. Entries in light grey do not appear in
  our examples, but are part of our testing. Dark grey entries are not tested.}
\label{tab:psi}
\end{table}

\paragraph*{Views and Security Classes}

The security context consists of a stack of \emph{views}, where a view is a
function mapping
each state element to a {\it security class}---one of
\(\public\), \(\unsealed\), \(\object\), or \(\sealed\).

State elements that are outside of the stack---general-purpose memory used for
globals and the heap, as well as the code region and globally shared
registers---are always labeled \(\public\). We place security requirements on some
\(\public\) elements for purposes of the well-bracketed control flow \(\wbcf\) property, and a
given enforcement mechanism
might restrict their access (e.g., by rendering code immutable), but for integrity
and confidentiality purposes they are considered accessible at all times.

When a function is newly activated, every stack location that is available
for use but not yet initialized
is \(\unsealed\). From the perspective of the caller, the callee has no obligations
regarding its use of free elements.

Arguments are marked \(\object\), meaning that their contents may be
used safely.
When a function allocates memory for its own stack frame, that memory will also be \(\object\).
Then, on a call, \(\object\) elements that are not being used to communicate with
the callee will become \(\sealed\)---i.e., reserved for an inactive principal
and expected to be unchanged when it becomes active again.

\paragraph*{Instantiating the Framework}

Conceptually, the following steps are needed to instantiate the framework to a specific machine
and coding conventions: (i) define the base machine semantics, including any hardware
security enforcement features; (ii) identify the set of
security-relevant operations and rules required by the coding conventions; (iii) determine
how to label machine instructions with security-relevant
operations as appropriate; (iv) specify the form of observable events.

\paragraph*{Threat Model and Limitations}

When our properties are used to evaluate a system, the threat model
will depend on the details of that system. However, there are some
constraints that our design puts on any system. In particular, we must
trust that the security-relevant operations have been correctly labeled.
If a compiled function
call is not marked as such, then the caller's data might not be
protected from the callee; conversely, marking too many operations as
calls may cause otherwise safe programs to be rejected.

We do not assume that low-level code adheres to any single calling
convention or is being used to implement any particular
source-language constructs.
Indeed, if the source language is C, then high-level programs might
contain undefined behavior, in which case they might be compiled to
arbitrary machine code.


In general, it is impossible to distinguish buggy machine code from an
attacker.  In examples, we often identify one function or another as
an attacker, but our framework does not require any static division between trusted
and untrusted code, and we aim to protect even buggy code.

This is a strong threat model, but it does omit some important aspects
of stack safety in real systems: in particular, it does not address
concurrency.  Hardware and timing attacks are also out of scope.

\section{Properties by Example}
\label{sec:example}

In this section, we introduce our security properties by means
of small code examples, using a simple set of security-relevant operations for
calls, returns, and private allocations.

\Cref{fig:main} gives C code and possible corresponding compiled 64-bit RISC-V code
for a function {\tt main}, which
takes an argument {\tt secret} and initializes a local variable {\tt sensitive} to contain
potentially sensitive data.
Then {\tt main} calls another function {\tt f},
and afterward it performs a test on {\tt sensitive} to decide whether
to output {\tt secret}.  Since {\tt sensitive} is initialized to 0,
the test should always fail, and {\tt main} should instead output the return value of {\tt f}.
Output is performed by writing to the special global {\tt out},
and we assume that such writes are the only observable events in the system.

The C code is compiled using the standard RISC-V calling conventions~\cite{RISC-V-CC}.
In particular, the function's first argument and its
return value are both passed in {\tt a0}.
Memory is byte addressed, and the stack grows towards
lower addresses. We assume that {\tt main} begins at address 0 and its
callee {\tt f} at address 100. The annotations in the right-hand column are
security-relevant operations, described further below.
The assembly is a simplified but otherwise typical compilation of the
source code into RISC-V; its details are less important than the positions
of the security-relevant operations.

\begin{figure}
  \newcommand{\figonebox}[1][]{\genbox{30pt}{blue!20}{#1}}
  ~~~~~~\begin{subfigure}{\columnwidth}
    {\tt
      volatile int out;

      void main(int secret) \{

      ~ ~ int sensitive = 0;

      ~ ~ int res = f();

      ~ ~ if (sensitive == 42)

      ~ ~ ~ ~ out = secret;

      ~ ~ else

      ~ ~ ~ ~ out = res;

      \}}
  \end{subfigure}
  \begin{subfigure}{\columnwidth}
    \begin{tabular}{r l | l}
      \labeledrow{0:}{addi sp,sp,-20}{\(\mathbf{alloc} ~ (-20,20)\)}
      \labeledrow{4:}{sd ra,12(sp)}{}
      \labeledrow{8:}{sw a0,8(sp)}{}
      \labeledrow{12:}{sw zero,4(sp)}{}
      \labeledrow{16:}{jal f,ra}{\(\mathbf{call} ~ \emplist \)}
      \labeledrow{20:}{sw a0,0(sp)}{}
      \labeledrow{24:}{lw a4,4(sp)}{}
      \labeledrow{28:}{li a5,42}{}
      \labeledrow{32:}{bne a4,a5,L1}{}
      \labeledrow{36:}{lw a0,8(sp)}{}
      \labeledrow{40:}{sw a0,out}{}
      \labeledrow{44:}{j L2}{}
      \labeledrow{L1, 48:}{lw a0,0(sp)}{}
      \labeledrow{52:}{sw a0,out}{}
      \labeledrow{L2, 56:}{ld ra,12(sp)}{}
      \labeledrow{60:}{addi sp,sp,20}{\(\mathbf{dealloc} ~ (0,20)\)}
      \labeledrow{64:}{jalr ra}{\(\mathbf{return}\)}
    \end{tabular}
  \end{subfigure}
  \begin{subfigure}{\columnwidth}
    \center
    \vspace{\abovedisplayskip}
    \(\xleftarrow{\figonebox[\dots] \stackrel{\textsc{\normalsize sp}}{\figonebox[\tt res]}
    \stackrel{\textsc{\normalsize 4(sp)}}{\figonebox[\tt sens]}
    \stackrel{\textsc{\normalsize 8(sp)}}{\figonebox[\tt sec]}
    \stackrel{\textsc{\normalsize 12(sp)} \hfill}{\figonebox[\(\mathtt{ra_1}\)]\figonebox[\(\mathtt{ra_2}\)]}
    }\)
  \end{subfigure}

  \caption{Example: C and assembly code for {\tt main} and layout of its stack frame (the stack grows to the left).}
  \label{fig:main}
\end{figure}

Now, suppose that {\tt f} is actually an attacker seeking
to leak {\tt secret}. It might do so in a number of ways, shown as snippets of
assembly code in \cref{fig:f}.
Leakage is most obviously viewed as a violation of {\tt main}'s {\it confidentiality}.
In \cref{subfig:direct}, {\tt f} takes an offset from the stack
pointer, accesses {\tt secret}, and directly outputs it.  More
subtly, even if it is somehow prevented from outputting {\tt secret}
directly, {\tt f}
can instead return its value so that {\tt main} stores it to {\tt out},
as in \cref{subfig:indirect}.
Beyond simply reading {\tt secret}, the attacker might overwrite {\tt sensitive}
with 42, guaranteeing that {\tt main} publishes its own secret unintentionally
(\cref{subfig:integrity}); this does not violate {\tt main}'s
confidentiality, but
rather its {\it integrity}.
In \cref{subfig:WBCF}, the attacker arranges to return to the
wrong instruction, thereby bypassing the check and publishing {\tt
  secret} regardless; this
violates the program's {\it well-bracketed control flow} (\(\wbcf\)).
In \cref{subfig:WBCF2}, a different attack violates \(\wbcf\), this time
by returning to the correct program counter but with the wrong stack pointer.
(We pad some of these variants with {\tt nop}s just so that all the
snippets have the same length, which keeps the step numbering uniform in~\cref{fig:exec1}.)

\begin{figure}
  \begin{subfigure}[b]{\columnwidth}
    \vspace{\abovedisplayskip}
    \begin{tabular}{r l | l}
      \labeledrow{100:}{lw a4,8(sp)}{}
      \labeledrow{104:}{sw a4,out}{}
      \labeledrow{108:}{li a0,1}{}
      \labeledrow{112:}{jalr ra}{\(\mathbf{return}\)}
    \end{tabular}
    \caption{Leaking {\tt secret} directly}
    \label{subfig:direct}
  \end{subfigure}
  \begin{subfigure}[b]{\columnwidth}
    \vspace{\abovedisplayskip}
    \begin{tabular}{r l | l}
      \labeledrow{100:}{lw a4,8(sp)}{}
      \labeledrow{104:}{mov a0,a4}{}
      \labeledrow{108:}{nop}{}
      \labeledrow{112:}{jalr ra}{\(\mathbf{return}\)}
    \end{tabular}
    \caption{Leaking {\tt secret} indirectly}
    \label{subfig:indirect}
  \end{subfigure}
  \begin{subfigure}[b]{\columnwidth}
    \vspace{\abovedisplayskip}
    \begin{tabular}{r l | l}
      \labeledrow{100:}{li a5,42}{}
      \labeledrow{104:}{sw a5,4(sp)}{}
      \labeledrow{108:}{li a0,1}{}
      \labeledrow{112:}{jalr ra}{\(\mathbf{return}\)}
    \end{tabular}
    \subcaption{Attacking {\tt sensitive}}
    \label{subfig:integrity}
  \end{subfigure}
  \begin{subfigure}[b]{\columnwidth}
    \vspace{\abovedisplayskip}
    \begin{tabular}{r l | l}
      \labeledrow{100:}{addi ra,ra,16}{}
      \labeledrow{104:}{nop}{}
      \labeledrow{108:}{nop}{}
      \labeledrow{112:}{jalr ra}{\(\mathbf{return}\)}
    \end{tabular}
    \subcaption{Attacking control flow}
    \label{subfig:WBCF}
  \end{subfigure}
  \begin{subfigure}[b]{\columnwidth}
    \vspace{\abovedisplayskip}
    \begin{tabular}{r l | l}
      \labeledrow{100:}{addi sp,sp,8}{}
      \labeledrow{104:}{nop}{}
      \labeledrow{108:}{nop}{}
      \labeledrow{112:}{jalr ra}{\(\mathbf{return}\)}
    \end{tabular}
    \subcaption{Attacking stack pointer integrity}
    \label{subfig:WBCF2}
  \end{subfigure}

  \caption{Example: assembly code alternatives for {\tt f} as an attacker.}
  \label{fig:f}
\end{figure}

The security semantics for this program is based
on the security-relevant events noted in the right columns of \cref{fig:main,fig:f},
namely execution of instructions that allocate or deallocate space (specified by
an \(\SP\)-relative offset and size), make a call (with a specified list of argument registers),
or make a return.

Our security semantics attaches a security context to the machine state,
consisting of a view \(V\) and a stack \(\sigma\) of pending activations' views.
\Cref{fig:exec1} shows how the security context evolves over the first few
steps of the program.  (The formal details of the security semantics are described in
\cref{sec:formal}, and the context evolution rules are formalized in \cref{fig:basicops}.)
Execution begins at the start of {\tt main}, with the program counter
(\(\PCname\)) set to zero and the stack pointer (\(\SP\)) at address 1000.
State transitions are numbered and may be labeled with a security operation, written
\(\downarrow \psi\), between steps.

The initial view \(V_0\) maps all stack addresses below \(\SP\) to \(\unsealed\) and the remainder of
memory to \(\public\). The sole used argument register, {\tt a0}, is mapped to \(\object\);
other caller-save registers are mapped to \(\unsealed\) and callee-save registers to \(\sealed\).
Step 1 allocates a word each for {\tt secret}, {\tt sensitive}, and {\tt res}, as well
as two words for the return address; this has the
effect of marking those bytes \(\object\).
We use \(V\llbracket\ldots\rrbracket\) to denote updates to \(V\).

\begin{figure*}
  \begin{tabular}{|r|r||l|r}
    \cline{1-3}
    \(\PCname\) & \(\SP\) & Context &
    \multirow{3}{*}{\(\underbrace{\dots \freebox \freebox \freebox \freebox \freebox
        \freebox \freebox \freebox \freebox \freebox}_\unsealed
      \! \underbrace{\stackrel{\stackrel{\SP}{\downarrow}}{\pubbox} \!\! \pubbox \pubbox \dots}_\public
      ~ \stackrel{\mathtt{a0}}{\objbox\objbox} ~ \stackrel{\mathtt{a4}}{\freebox\freebox}
      ~ \stackrel{\mathtt{a5}}{\freebox\freebox}
      \)} \\
    \cline{1-3}
    0 & 1000 & \(V_0, \emplist\)
    \\
    \cline{1-3}
    \multicolumn{3}{l}{\multirow{2}{*}{\(1 \Big\downarrow \mathbf{alloc} ~ (-20,20)\)}} & \\
    \multicolumn{3}{l}{} &
    \multirow{3}{*}{\(\underbrace{\dots \freebox \freebox \freebox \freebox \freebox}_\unsealed
      \! \underbrace{\stackrel{\stackrel{\SP}{\downarrow}}{\objbox} \!\! \objbox \objbox \objbox \objbox}_\object
      \! \underbrace{\pubbox \pubbox \pubbox \dots}_\public
      ~ \stackrel{\mathtt{a0}}{\objbox\objbox} ~ \stackrel{\mathtt{a4}}{\freebox\freebox}
      ~ \stackrel{\mathtt{a5}}{\freebox\freebox}
      \)}
    \\
    \cline{1-3}
    4 & 980 & \(V_1 = V_0 \llbracket 980..999 \mapsto \object\rrbracket, \emplist\) &
    \\
    \cline{1-3}
    \multicolumn{3}{l}{\multirow{2}{*}{2-4 \(\Big\downarrow\)}} \\ \multicolumn{3}{l}{} \\
    \cline{1-3}
    16 & 980 & \(V_1, \emplist\) & \\
    \cline{1-3}
    \multicolumn{3}{l}{\multirow{2}{*}{\(5 \Big\downarrow \mathbf{call} ~ 100 ~ \emplist\)}} & \\
    \multicolumn{3}{l}{} &
    \multirow{3}{*}{\(\underbrace{\dots \freebox \freebox \freebox \freebox \freebox}_\unsealed
      \! \underbrace{\stackrel{\stackrel{\SP}{\downarrow}}{\sealbox} \!\! \sealbox \sealbox \sealbox \sealbox}_\sealed
      \! \underbrace{\pubbox \pubbox \pubbox \dots}_\public
      ~ \stackrel{\mathtt{a0}}{\freebox\freebox} ~ \stackrel{\mathtt{a4}}{\freebox\freebox}
      ~ \stackrel{\mathtt{a5}}{\freebox\freebox}
      \)}
    \\
    \cline{1-3}
    100 & 980 & \(V_2 = V_1 \llbracket 980..999 \mapsto \sealed, \mathtt{a0} \mapsto \unsealed\rrbracket,[V_1]\) & \\
    \cline{1-3}
    \multicolumn{3}{l}{\multirow{2}{*}{6-8 \(\Big\downarrow\)}} \\ \multicolumn{3}{l}{} \\
    \cline{1-3}
    112 & 980 & \(V_2,[V_1]\) \\
    \cline{1-3}
    \multicolumn{3}{l}{\multirow{2}{*}{\(9 \Big\downarrow \mathbf{return}\)}} & \\
    \multicolumn{3}{l}{} & \multirow{3}{*}{\(\underbrace{\dots \freebox \freebox \freebox \freebox \freebox}_\unsealed
      \! \underbrace{\stackrel{\stackrel{\SP}{\downarrow}}{\objbox} \!\! \objbox \objbox \objbox \objbox}_\object
      \! \underbrace{\pubbox \pubbox \pubbox \dots}_\public
      ~ \stackrel{\mathtt{a0}}{\objbox\objbox} ~ \stackrel{\mathtt{a4}}{\freebox\freebox}
      ~ \stackrel{\mathtt{a5}}{\freebox\freebox}
      \)}
    \\
    \cline{1-3}
    20 & 980  & \(V_1, \emplist\) &
    \\
    \cline{1-3}
    \multicolumn{2}{l}{} \\
  \end{tabular}
  \caption{Execution of example up through the return from {\tt f}. In stack diagrams, addresses increase to the right, stack grows to the left, and boxes represent 4-byte words.}
\label{fig:exec1}
\end{figure*}
At step 5, the current principal's record is pushed onto the inactive list.
The callee's view is updated from the caller's such that all \(\object\) memory locations
become \(\sealed\). (For now we assume no sharing of stack memory between activations; data is
passed only through argument registers, which remain active. In the presence of memory
sharing, some memory would remain active, too.)
Function {\tt f} does not take any arguments; if it did, any registers containing them would be
mapped to \(\object\), while any non-argument, caller-saved
registers are mapped to \(\unsealed\). In the current example, only register {\tt a0}  changes
security class. All callee-save registers remain \(\sealed\) for all calls, so
if, in the example, we varied the assembly code for {\tt main} so that {\tt sensitive} was stored
in a callee-save register (e.g., {\tt s0}) rather than in memory, its security class would still
be \(\sealed\) at the entry to {\tt f}.
At step 9, {\tt f} returns and the topmost inactive view, that of {\tt main}, is restored.

We now show how this security semantics can be used to define notions of confidentiality,
integrity, and correct control flow in such a way that many classes of
bad behavior, including the attacks in \cref{fig:f}, are
detected as security violations.

\paragraph*{Well-Bracketed Control Flow}

To begin with, if {\tt f} returns to an unexpected place (i.e., \(\PCname \neq 20\) or
\(\SP \neq 980\)), we say that it has violated \(\wbcf\). \(\wbcf\) is a relationship between
call steps and their corresponding return steps: just after the return, the program
counter should be at the next instruction below the call,
and the stack pointer should have the same value that it had before the call.
Both of these are essential for security. In \cref{subfig:WBCF}, the attacker adds
16 to the return address and then returns; this bypasses the {\tt if}-test in the code and outputs
{\tt secret}.
In \cref{subfig:WBCF2}, the attacker returns with \(\SP' = 988\) instead of the
correct \(\SP = 980\). In this scenario, given the layout of {\tt main}'s frame,
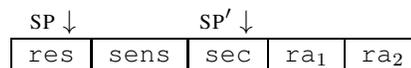
\begin{center}
\begin{tabular}{| l | l | l | l | l |}
  \multicolumn{1}{r}{\(\SP \downarrow\)} &
  \multicolumn{2}{r}{\(\SP' \downarrow\)} \\
  \hline
  {\tt res} & {\tt sens} & {\tt sec} & \(\mbox{\tt ra}_1\) & \(\mbox{\tt ra}_2\) \\
  \hline
\end{tabular}
\end{center}

\vspace{\abovedisplayskip}

\noindent
{\tt main}'s attempt to read {\tt sensitive} may instead
read part of the return address, and its attempt to output
{\tt res} will instead output {\tt secret}.

Before the call, the program counter is 16 and the stack pointer is 980.
So we define a predicate on states that should hold just after the return:
\(\ret\ \mach \triangleq \mach[\PCname] = 20 \wedge \mach[\SP] = 980\).
We can identify the point just after the return (if a return occurs)
as the first state in which the pending call stack is smaller than it was
just after the call.
\(\wbcf\) requires that, if \(\mach\) is the state at that point, then \(\ret ~ \mach\) holds.
This property is formalized in \cref{tab:props}, line 1.


\paragraph*{Stack Integrity}

Like \(\wbcf\), stack integrity defines a condition at the call that must hold upon
return. This time the condition applies to all of the memory in the caller's
frame. In \cref{fig:exec1} we see the lifecycle of an allocated frame:
upon allocation, the view labels it \(\object\), and when a call is made, it instead
becomes \(\sealed\). Intuitively, the integrity of {\tt main}
is preserved if, when control returns to it, any \(\sealed\) elements
are identical to when it made the call.
Again, we need to know when a caller has been returned to,
and we use the same mechanism of checking the depth of the call stack.
In the case of the call from {\tt main} to {\tt f}, the \(\sealed\) elements are the
addresses 980 through 999 and callee-saved registers such as
the stack pointer. Note that callee-saved registers often change
during the call---but if the caller accesses them after the call, it should find them
restored to their prior value.

While it would be simple to define integrity as ``all sealed elements retain their
values after the call,'' this would be stricter than necessary. Suppose that
a callee overwrites some data of its caller, but the caller never accesses that data
(or only does so after re-initializing it). This would be harmless, with the callee
essentially using the caller's memory as scratch space, but the caller never seeing any change.

For a set of elements \(\components\),
a pair of states \(\mach\) and \(\nach\) are {\em \(\components\)-variants} if
their values an only disagree on elements in \(\components\).
We say that the elements of \(\components\) are \emph{irrelevant}
in \(\mach\) if they can be replaced by arbitrary other values without changing the
observable behavior of the machine. All other elements are \emph{relevant}.\footnote{
This story is slightly over-simplified. If an enforcement mechanism maintains
additional state associated with elements, such as tags, we don't want that
state to vary. This is touched on in \cref{sec:props}.
}

We define \emph{caller integrity} (\(\clri\))  as the property that
every relevant element that is \(\sealed\) under the callee's view is restored
to its original value at the return point.
(This property is formalized in \cref{tab:props}, line 2).

\newcommand{\figfourbox}[1][]{\genbox{20pt}{red}{#1}}

\begin{figure}
  \centering
  \[
  \stackrel{\texttt{res}}{\figfourbox[0]}
  \stackrel{\texttt{sens}}{\figfourbox[0]}
  \stackrel{\texttt{sec}}{\figfourbox[5]}
  \stackrel{\texttt{ra}}{\figfourbox[0]\figfourbox[0]}\]
  \[\big\Downarrow\]
  \[
  \stackrel{\texttt{res}}{\figfourbox[0]}
  \stackrel{\texttt{sens}}{\genbox{20pt}{red!50}{\bf 42}}
  \stackrel{\texttt{sec}}{\figfourbox[5]}
  \stackrel{\texttt{ra}}{\figfourbox[0]\figfourbox[0]}\]
  \[\overbrace{
    \stackrel{\texttt{res}}{\figfourbox[0]}
    \stackrel{\texttt{sens}}{\genbox{20pt}{\leftvariant}{\bf 42}}
    \stackrel{\texttt{sec}}{\figfourbox[5]}
    \stackrel{\texttt{ra}}{\figfourbox[0]\figfourbox[0]}
    \hspace{1cm}
    \stackrel{\texttt{res}}{\figfourbox[0]}
    \stackrel{\texttt{sens}}{\genbox{20pt}{\rightvariant}{\bf 0}}
    \stackrel{\texttt{sec}}{\figfourbox[5]}
    \stackrel{\texttt{ra}}{\figfourbox[0]\figfourbox[0]}}\]
  \[\stackrel{\hookrightarrow \mathtt{out}}{\genbox{20pt}{white}{5}} \hspace{0.5cm}
  \hspace{0.5cm}
  \stackrel{\hookrightarrow \mathtt{out}}{\genbox{20pt}{white}{1}}\]
  \caption{Integrity Violation: {\tt sensitive} changed, and if varied, changes future outputs}
  \label{fig:variant}
\end{figure}

In our example setting, the observation trace consists of the sequence
of values written to {\tt out}. In \cref{subfig:integrity} the states before
and after the call differ in the value of {\tt sensitive}. \Cref{fig:variant}
shows the states before and after the call, which disagree on the value at
{\tt sensitive}. If we consider a
variant of the original return state in which {\tt sensitive} is 0 (orange)
as opposed to 42 (blue), that state will eventually output 1, while the actual
execution outputs 5. This means that {\tt sensitive} is relevant.

To be more explicit, similar to \(\wbcf\), we define
\(\intProp\) as a predicate on states that holds if
all relevant sealed addresses in \(\mach\) are the same as after step 5.
We require that \(\intProp\) hold on the state following the matching return,
which is reached by step 9. Here {\tt sensitive} has obviously changed, but we
just saw that it is relevant.

\paragraph*{Caller Confidentiality}

We treat confidentiality as a form of non-interference as well: the confidentiality of a caller
means that its callee's behavior is dependent only on publicly visible data,
not the caller's private state. This also requires that the callee initialize
memory before reading it.
As we saw in the examples, we must consider both the observable events
that the callee produces during the call and the changes that the callee makes to the state that might
affect the caller after the callee returns.

Consider the state after step 5, shown at the top of \cref{fig:variant2},
with the attacker code from \cref{subfig:direct} and the assumption that
{\tt secret} has the value 5. We take a variant state over
the set of elements that are \(\sealed\) in \(V_2\) (orange), and compare it to the
original (blue). During the execution, the value of {\tt secret} is written
to the output, and the information leak is evidenced by the fact that the
outputs do not agree---the original outputs 5, while the variant outputs 3.
This is a violation of
{\it internal confidentiality} (formalized in \cref{tab:props}, line 3a).

\newcommand{\leftbox}[1][]{\genbox{20pt}{\leftvariant}{#1}}
\newcommand{\rightbox}[1][]{\genbox{20pt}{\rightvariant}{#1}}

\begin{figure}
    \centering
    \[
    \stackrel{\texttt{res}}{\figfourbox[0]}
    \stackrel{\texttt{sens}}{\figfourbox[0]}
    \stackrel{\texttt{sec}}{\figfourbox[5]}
    \stackrel{\texttt{ra}}{\figfourbox[0]\figfourbox[0]}\]
    \[\overbrace{
    \stackrel{\texttt{res}}{\leftbox[0]}
    \stackrel{\texttt{sens}}{\leftbox[0]}
    \stackrel{\texttt{sec}}{\leftbox[5]}
    \stackrel{\texttt{ra}}{\leftbox[0]\leftbox[0]}
    \hspace{1cm}
    \stackrel{\texttt{res}}{\rightbox[1]}
    \stackrel{\texttt{sens}}{\rightbox[2]}
    \stackrel{\texttt{sec}}{\rightbox[3]}
    \stackrel{\texttt{ra}}{\rightbox[4]\rightbox[5]}}
    \]
    \[\raisebox{.5\height}{\Bigg\Downarrow} \hspace{1cm}
    \stackrel{\hookrightarrow \mathtt{out}}{\genbox{20pt}{white}{5}} \hspace{0.5cm}
    \raisebox{\height}{\(\not\approx\)}
    \hspace{0.5cm}
    \stackrel{\hookrightarrow \mathtt{out}}{\genbox{20pt}{white}{3}} \hspace{1cm}
    \raisebox{.5\height}{\Bigg\Downarrow}
    \]
    \[
    \stackrel{\texttt{res}}{\leftbox[0]}
    \stackrel{\texttt{sens}}{\leftbox[0]}
    \stackrel{\texttt{sec}}{\leftbox[5]}
    \stackrel{\texttt{ra}}{\leftbox[0]\leftbox[0]}
    \hspace{1cm}
    \stackrel{\texttt{res}}{\rightbox[1]}
    \stackrel{\texttt{sens}}{\rightbox[2]}
    \stackrel{\texttt{sec}}{\rightbox[3]}
    \stackrel{\texttt{ra}}{\rightbox[4]\rightbox[5]}
    \]

  \caption{Internal Confidentiality Violation}
  \label{fig:variant2}
\end{figure}

But, in \cref{subfig:indirect}, we also saw an attacker that exfiltrated the secret
by reading it and then returning it, in a context where the caller would output the returned
value. \Cref{fig:variant3} shows the behavior of the same variants under this attacker,
but in this case, there is no output during the call. Instead the value of {\tt secret} is
extracted and placed in {\tt a0}, the return value register.
\begin{figure}
    \centering
    \[
    \stackrel{\texttt{res}}{\figfourbox[0]}
    \stackrel{\texttt{sens}}{\figfourbox[0]}
    \stackrel{\texttt{sec}}{\figfourbox[5]}
    \stackrel{\texttt{ra}}{\figfourbox[0]\figfourbox[0]}\]
    \[\overbrace{
    \stackrel{\texttt{res}}{\leftbox[0]}
    \stackrel{\texttt{sens}}{\leftbox[0]}
    \stackrel{\texttt{sec}}{\leftbox[5]}
    \stackrel{\texttt{ra}}{\leftbox[0]\leftbox[0]}
    \hspace{1cm}
    \stackrel{\texttt{res}}{\rightbox[1]}
    \stackrel{\texttt{sens}}{\rightbox[2]}
    \stackrel{\texttt{sec}}{\rightbox[3]}
    \stackrel{\texttt{ra}}{\rightbox[4]\rightbox[5]}}
    \]
    \[\raisebox{.5\height}{\Bigg\Downarrow} \hspace{1cm} \stackrel{\mathtt{a0}}{\leftbox[\bf 0]} \hspace{1cm}
    \stackrel{\mathtt{a0}}{\rightbox[\bf 6]} \hspace{1cm} \raisebox{.5\height}{\Bigg\Downarrow}\]
    \[
    \stackrel{\texttt{res}}{\leftbox[0]}
    \stackrel{\texttt{sens}}{\leftbox[0]}
    \stackrel{\texttt{sec}}{\leftbox[5]}
    \stackrel{\texttt{ra}}{\leftbox[0]\leftbox[0]}
    \hspace{1cm}
    \stackrel{\texttt{res}}{\rightbox[1]}
    \stackrel{\texttt{sens}}{\rightbox[2]}
    \stackrel{\texttt{sec}}{\rightbox[3]}
    \stackrel{\texttt{ra}}{\rightbox[4]\rightbox[5]}
    \]
    \[\stackrel{\mathtt{a0}}{\leftbox[\bf 5]} \hspace{1cm}
    \stackrel{\mathtt{a0}}{\rightbox[\bf 3]}\]
  \caption{Return-time Confidentiality Violation}
  \label{fig:variant3}
\end{figure}

At the end of the call, we can deduce that every element on which
the variant states disagree must carry some information derived from
the original varied elements. In most cases, that is because the element
is one of the original varied elements and has not changed during the
call, which does not represent a leak. But in the case of {\tt a0}, it
has changed during the call, {\em and} the return states do not agree
on its value. This represents data that has been leaked, and should
not be used to affect future execution.
Unless {\tt a0} happens to be irrelevant to the caller, this example
is a violation of what we term {\it return-time confidentiality}
(formalized in \cref{tab:props}, line 3b).

Structurally, return-time confidentiality resembles integrity, but now dealing with
variants. We begin with a state immediately following
a call, \(\mach\). We consider an arbitrary variant state,
\(\nach\), which may vary any element that is \(\sealed\) or \(\unsealed\),
i.e., any element that is not used legitimately to pass arguments. Caller confidentiality
therefore can be thought of as the callee's insensitivity to elements in its initial state
that are not part of the caller-callee interface.

We define a binary relation \(\confProp\) on pairs of states,
which holds on eventual return states \(\mach'\) and \(\nach'\)
if all relevant elements are {\em uncorrupted} relative to \(\mach\) and \(\nach\).
An element is {\em corrupted} if it differs between \(\mach'\) and \(\nach'\),
and it either changed between \(\mach\) and \(\mach'\) or between \(\nach\) and \(\nach'\).

Finally, we define \emph{caller confidentiality} (\(\clrc\)) as the
combination of internal and return-time confidentiality (\cref{tab:props}, line 3).

\paragraph*{The Callee's Perspective}

We presented our initial example from the perspective of the caller, but a callee
may also have privilege that its caller lacks, and which must be protected from the
caller. Consider a function that makes a privileged system call to obtain a secret key,
and uses that key to perform a specific task. An untrustworthy or erroneous caller might
attempt to read the key out of the callee's memory after return, or to influence the callee
to cause it to misuse the key itself!

Where the caller's confidentiality and integrity are concerned with protecting specific,
identifiable state---the caller's stack frame---their callee equivalents are concerned
with enforcing the expected interface between caller and callee. Communication between
the principals should occur only through the state elements that are designated for the
purpose: those labeled \(\public\) and \(\object\).

Applying this intuition using our framework, \emph{callee confidentiality} (\(\clec\))
turns out to resemble \(\clri\), extended to every element that is not marked \(\object\)
or \(\public\) at call-time. The callee's internal behavior is represented by those
elements that change over the course of its execution, and which are not part of the
interface with the caller. At return, those elements should become irrelevant to the
subsequent behavior of the caller.

Similarly, in \emph{callee integrity} (\(\clei\)), only elements marked \(\object\)
or \(\public\) at the call should influence the behavior of the callee. It may seem
odd to call this integrity, as the callee does not have a private state. But
an erroneous callee that performs a read-before-write within its stack
frame, or which uses a non-argument register without initializing it, is vulnerable
to its caller seeding those elements with values that will change its behavior.
The fact that well-behaved callees have integrity by definition is probably why
callee integrity is not typically discussed.

\section{Formalization}
\label{sec:formal}

We now give a formal description of our machine model, security semantics,
and properties. Our definitions abstract over: (i) the details of  the target machine
architecture and ABI, (ii) the set of security-relevant operations and their effects on
the security context, (iii) the set of observable events, and (iv) a notion of value compatibility.

\subsection{Machine}
The building blocks of a machine are {\em words} and {\em registers}.
Words are ranged over by \(\word\) and, when used as addresses, \(\addr\),
and are drawn from the set \(\WORDS\).
Registers in the set \(\REGS\) are ranged over by \(\reg\), with the stack pointer
given the special name \(\SP\);
some registers may be classified as caller-saved (CLR) or callee-saved (CLE).
Along with the program counter, \(\PCname\), these are referred to as
{\em state elements} \(\component\) in the set \(\COMPONENTS ::= \PCname | \WORDS | \REGS\).

A {\em machine state} \(\mach \in \MACHS\) is a map from state elements to a set \(\mathcal{V}\) of
\emph{values}.
Each value \(v\) contains a \emph{payload} word, written \(|v|\).
We write \(\mach[\component]\) to denote the value of \(\mach\) at
\(\component\)  and \(\mach[v]\) as shorthand for \(\mach[|v|]\).
Depending on the specific machine being modeled, values may also contain other
information relevant to hardware enforcement (such as a tag).
When constructing variants (see~\cref{sec:props}, this additional information should
not be varied. To capture this idea, we assume a given \emph{compatibility} equivalence relation \(\sim\) on values,
and lift it element-wise to states.  Two values should be compatible if their
non-payload information (e.g., their tag) is identical.

The machine has a step function \(\mach \xrightarrow{\bar{\psi},\obs} \mach'\).
Except for the annotations over the arrow, this function just encodes the usual
ISA description of the machine's instruction set. The annotations serve to connect
the machine's operation to our security setting:
\(\bar{\psi}\) is a list of security-relevant operations drawn from an assumed given set \(\Psi\),
and \(\obs\) is an (potentially silent) observable event; these are described further below.

\subsection{Security semantics}

The security semantics operates in parallel with the machine.
Each state element (memory word or register) is given a \emph{security class}
\(l \in \{\public, \object, \sealed, \unsealed\}\).
A \emph{view} \(V \in \mathit{VIEW}\) maps elements to security classes.
For any security class \(l\), we write \(l(V)\)
to denote the set of elements \(\component\) such that \(V ~ \component = l\).
The {\it initial view} \(V_0\) maps all stack locations to \(\unsealed\),
all other locations to \(\public\), and registers based on which set they
belong to: \(\sealed\) for callee-saved, \(\unsealed\) for caller-saved except for those
that contain arguments at the start of execution, which are \(\object\), and \(\public\) otherwise.

A (security) \emph{context} is
a pair of the current activation's view and
a list of views representing the call stack (pending inactive
principals), ranged over by \(\sigma\).
\[\context \in \CONTEXTS ::= \mathit{VIEW \times list ~ VIEW}\]
The initial context is \(\context_0 = (V_0, \emplist)\).

\Cref{sec:example} describes informally how the security context evolves as the system performs
security-relevant operations. Formally, we combine each machine state with a context
to create a {\it combined state} \(s = (\mach,\context)\) and lift the transition
to \(\stepstounder{}\) on combined states.
At each step, the context updates based on an assumed given function
\(Op : \MACHS \rightarrow \CONTEXTS \rightarrow \Psi \rightarrow \CONTEXTS\).
Since a single step might correspond to multiple operations, we apply
\(Op\) as many times as needed, using \(\mathit{foldl}\).

\judgmenttwo{\(\mach \xrightarrow{\overline{\psi},\obs} \mach' \)}
            {\(\mathit{foldl} ~ (Op ~ \mach) ~ \context ~ \overline{\psi} = \context'\)}
            {\((\mach,\context) \stepstounder{\overline{\psi},\obs} (\mach', \context')\)}

A definition of \(Op\) is most convenient to present decomposed into
rules for each operation. We have already seen the intuition behind the rules for
\(\mathbf{alloc}\), \(\mathbf{call}\), and \(\mathbf{ret}\).
For the machine described in the example, the \(Op\) rules would be those
found in \cref{fig:basicops}.
Note that \(Op\) takes as its first argument the state {\it before} the step.

\begin{figure}
    \[\mathit{range} ~ \reg ~ \mathit{off} ~ \mathit{sz} ~ \mach \triangleq
    \{\mach[\reg]+i | \mathit{off} \leq i < \mathit{off+sz}\}\]

    \judgmentbr[~Alloc]
               {\(\components = \mathit{range} ~ \SP ~ \mathit{off} ~ \mathit{sz} ~ \mach \cap \unsealed(V)\)}
               {\(V' = V \llbracket \addr \mapsto \object \mid \addr \in \components \rrbracket\)}
               {\(Op ~ \mach ~ (\mathbf{alloc} ~ \mathit{off, sz}) ~ (V,\sigma) = (V',\sigma)\)}
               
    \vspace{\abovedisplayskip}

    \judgmentbr[~Dealloc]
               {\(\components = \mathit{range}~ \SP ~ \mathit{off} ~ \mathit {sz} ~ \mach \cap \object(V)\)}
               {\(V' = V \llbracket \addr \mapsto \unsealed \mid \addr \in \components \rrbracket\)}
               {\(Op ~ \mach ~ (\mathbf{dealloc} ~ \mathit{off, sz}) ~ (V,\sigma) = (V',\sigma)\)}

    \vspace{\abovedisplayskip}
    
    \judgment[~Call]
             {\(V' = \lambda \component .
               \begin{cases}
                 \unsealed & \textnormal{if } k \in CLR \\
                 \public & \textnormal{if } k \in \overline{\reg_{\mathit{args}}} \\
                 \sealed & \textnormal{if } k \in \WORDS \textnormal{ and } k \in \object(V) \\
                 V(k) & \textnormal{otherwise} \\
               \end{cases}\)}
             {\(Op ~ \mach ~ (\mathbf{call} ~ \addr_{target} ~ \overline{\reg_{args}})
               ~ (V,\sigma) = (V',V::\sigma)\)}

    \vspace{\abovedisplayskip}             
    
    \judgment[~Return]
             {}
             {\(Op ~ \mach ~ \mathbf{return} ~ (\_, (V,\sigma')) = (V, \sigma')\)}
             \caption{Basic Operations}
  \label{fig:basicops}
\end{figure}

\subsection{Events and Traces}
\label{sec:events}

We abstract over the events that can be observed in the system, assuming just
a given set \(\OBSS\) that contains at least the element \(\tau\), the silent
event. Other events might represent certain function calls (i.e., system calls)
or writes to special addresses representing memory-mapped regions.
A {\em trace} is a nonempty, finite or infinite sequence
of events, ranged over by \(\obsT\).
We use ``\(\notfinished{}{}\)'' to represent ``cons'' for traces, reserving ``::''
for list-cons.

We are particularly interested in traces that end just after a function returns.
We define these in terms of the depth \(d\) of the security context's call stack \(\sigma\).
We write \(d \hookrightarrow s\) for the trace of execution from a state \(s\)
up to the first point where the stack depth is smaller than \(d\), defined
coinductively by these rules:

\judgment[~Done]
         {\(|\sigma| < d\)}
         {\(d \hookrightarrow (\mach,(V,\sigma)) = \tau\)}

\vspace{\abovedisplayskip}
         
\judgmenttwobrlong[~Step]
                  {\(|\sigma| \geq d\)}
                  {\(d \hookrightarrow (\mach',\context') = \obsT\)}
                  {\((\mach,(V,\sigma)) \stepstounder{\overline{\psi},\obs} (\mach',\context')\)}
                  {\(d \hookrightarrow (\mach,(V,\sigma)) = \notfinished{\obs}{\obsT}\)}

\noindent
When \(d = 0\), the trace will always be infinite because the machine never halts; in this case we
omit \(d\) and just write \(\hookrightarrow s\).

Two event traces $\obsT_1$ and $\obsT_2$ are {\em similar},
written \(\obsT_1 \eqsim \obsT_2\), if the sequence of non-silent events
is the same. That is, we compare up to deletion of \(\tau\) events.
Note that this results in an infinite silent trace being similar to
any trace. So, a trace that silently diverges due to a failstop will
be vacuously similar to all other traces.

\begin{minipage}{.4\columnwidth}
  \judgment[~SimRefl]{}{\(\obsT \eqsim \obsT\)}
\end{minipage}
\begin{minipage}{.4\columnwidth}
  \judgment[~SimEvent]
           {\(\obsT_1 \eqsim \obsT_2\)}
           {\(\notfinished{\obs}{\obsT_1} \eqsim \notfinished{\obs}{\obsT_2}\)}
\end{minipage}

\begin{minipage}{.4\columnwidth}
  \judgment[~SimLeft]
           {\(\obsT_1 \eqsim \obsT_2\)}
           {\(\notfinished{\tau}{\obsT_1} \eqsim \obsT_2\)}
\end{minipage}
\begin{minipage}{.5\columnwidth}
  \judgment[~SimRight]
           {\(\obsT_1 \eqsim \obsT_2\)}
           {\(\obsT_1 \eqsim \notfinished{\tau}{\obsT_2}\)}
\end{minipage}

\subsection{Variants, corrupted sets, and ``on-return'' assertions}
\label{sec:props}

Two (compatible) states are variants with respect to a set of elements \(\components\)
if they agree on the value of every element not in \(\components\).
Our notion of non-interference involves comparing the traces of such
\(\components\)-variants. We use this to define sets of irrelevant elements.
Recall that \(\sim\) is a policy-specific compatibility relation.

\definition The \emph{difference set} of two machine states \(\mach\) and \(\mach'\),
written \(\Delta(\mach,\mach')\),
is the set of elements \(\component\) such that \(\mach[\component] \not = \mach'[\component]\).

\definition Machine states \(\mach\) and \(\nach\) are {\em \(\components\)-variants},
written \(\mach \approx_\components \nach\), if \(\mach\sim\nach\) and
\(\Delta(\mach,\nach) \subseteq \components\).

\definition An element set \(\components\) is \emph{irrelevant} to state \((\mach,\context)\),
written \((\mach,\context) \parallel \components\), if for all
\(\nach\) such that \(\mach \approx_{\components} \nach\),
\(\hookrightarrow (\mach,\context) ~ \eqsim ~ \hookrightarrow (\nach,\context)\).

When comparing the behavior of variant states, we need a notion of how their
differences have influenced them.
\definition The {\em corrupted set} \(\bar{\Diamond}(\mach,\mach',\nach,\nach')\)
is the set \((\Delta(\mach,\mach') \cup \Delta(\nach,\nach')) \cap \Delta(\mach',\nach')\).

If we consider two execution sequences, one from \(\mach\) to \(\mach'\)
and the other from \(\nach\) to \(\nach'\),
then \(\bar{\Diamond}(\mach,\mach',\nach,\nach')\) is the set of elements that
change in one or both executions and end up with different values. Intuitively,
this captures the effect of any differences between \(\mach\) and \(\nach\), i.e.,
the set of values that are ``corrupted'' by those differences.

Our ``on-return'' assertions are defined using a second-order logical operator
\(d \uparrow P\), pronounced ``\(P\) holds on return from depth \(d\),''
where \(P\) is a predicate on machine states. This is a coinductive relation
similar to ``weak until'' in temporal logic---it also holds if the program never
returns from depth \(d\).

\vspace{\abovedisplayskip}

\judgmenttwo[~Returned]
            {\(|\sigma| < d\)}
            {\(P ~ \mach\)}
            {\((d \uparrow P) ~ (\mach, (V,\sigma))\)}

\vspace{\abovedisplayskip}            
            
\judgmenttwobrlong[~Step]
                  {\(|\sigma| \geq d\)}
                  {\((d \uparrow P) ~ (\mach', \context')\)}
                  {\((\mach, (V,\sigma)) \stepstounder{\overline{\psi},\obs} (\mach', \context')\)}
                  {\((d \uparrow P) ~ (\mach, (V,\sigma))\)}

Similarly, we give a analogous binary relation for use in confidentiality. We define \(\Uparrow\) so that
\((\mach,\context) ~ (d \Uparrow R) ~ (\mach',\context')\) holds if \(R\) holds on the
first states that return from depth \(d\) after \((\mach,\context)\) and \((\mach',\context')\),
respectively. Once again, \(\Uparrow\) is coinductive.

\vspace{\abovedisplayskip}
\judgmentthree[~Returned]
              {\(|\sigma_1| < d\)}
              {\(|\sigma_2| < d\)}
              {\(\mach_1 ~ R ~ \mach_2\)}
              {\((\mach_1,(V_1,\sigma_1)) ~ (d \Uparrow R) ~ (\mach_2,(V_2,\sigma_2))\)}

\judgmenttwobrlong[~Left]
                  {\(|\sigma_1| \geq d\)}
                  {\((\mach_1,(V_1,\sigma_1)) \stepstounder{\overline{\psi},\obs} (\mach_1',\context_1')\)}
                  {\((\mach_1',\context_1') ~ (d \Uparrow R) ~ (\mach_2,(V_2,\sigma_2))\)}
                  {\((\mach_1,(V_1,\sigma_1)) ~ (d \Uparrow R) ~ (\mach_2,(V_2,\sigma_2))\)}

\judgmenttwobrlong[~Right]
                  {\(|\sigma_2| \geq d\)}
                  {\((\mach_2,(V_2,\sigma_2)) \stepstounder{\overline{\psi},\obs} (\mach_2',\context_2')\)}
                  {\((\mach_1,(V_1,\sigma_1)) ~ (d \Uparrow R) ~ (\mach_2',\context_2')\)}
                  {\((\mach_1,(V_1,\sigma_1)) ~ (d \Uparrow R) ~ (\mach_2,(V_2,\sigma_2))\)}

\subsection{Properties}

\begin{table*}[h]
  \setlength{\tabcolsep}{1pt}
  \center
  \begin{tabular}{l r l l l}
    \rowcolor{black!20}
    1
    & \(\wbcf \triangleq\) & \((|\sigma'| \uparrow \ret) ~ (\mach', (V',\sigma'))\)
    & \(\textnormal{ where } \ret ~ \mach'' \triangleq \)
    \(\mach''[\SP] = \mach[\SP]\)
    & \(\textnormal{ for all calls } (\mach,(V,\sigma)) \stepstounder{} (\mach',(V',\sigma'))\) \\
    \rowcolor{black!20}
    & & & \(\textnormal{ \hspace{0.8in}} \land \mach''[\PCname] = \mach[\PCname]+\mathit{sz}\) & \textnormal{ where} \(\mathit{sz}\) is the size of instruction at \(\mach[\PCname]\) \\
    \rowcolor{black!10}
    2
    & \(\clri \triangleq\) & \((|\sigma| \uparrow \intProp) ~ (\mach,(V,\sigma))\)
    & \(\textnormal{ where } \intProp ~ \mach' \triangleq
    \mach' \parallel (\sealed(V) \cap \Delta(\mach,\mach'))\)
    & \(\textnormal{ for all call targets } (\mach,(V,\sigma))\) \\
    \rowcolor{black!20}
    3
    & \(\clrc \triangleq\) & \(\forall \nach \textnormal{ s.t. } \mach \approx_{\components} \nach,\)
    & \(\textnormal{ where } \components = \sealed(V)\)
    & \(\textnormal{ for all call targets } (\mach,(V,\sigma))\) \\
    \rowcolor{black!20}
    3a & & \(|\sigma| \hookrightarrow (\mach,(V,\sigma)) \simeq |\sigma| \hookrightarrow (\nach,(V,\sigma))\) & & \\
    \rowcolor{black!20}
    3b & & \(\textnormal{ and } (\mach,(V,\sigma)) ~ (|\sigma| \Uparrow \confProp) ~ (\nach,(V,\sigma))\)
    & \(\textnormal{ where } (\mach' ~ \confProp ~ \nach') \triangleq
    \mach' \parallel \bar{\Diamond}(\mach,\nach,\mach',\nach')\) & \\
    \rowcolor{black!10}
    4
    & \(\clec \triangleq\) & \((|\sigma| \uparrow \cconfProp) ~ (\mach,(V,\sigma))\)
    & \(\textnormal{ where } \cconfProp ~ \mach' \triangleq
    \mach' \parallel (\Delta(\mach,\mach') - \components)\)
    & \(\textnormal{ for all call targets } (\mach,(V,\sigma))\) \\
    \rowcolor{black!10}
    & & & \(\textnormal{ where } \components = \public(V) \cup \object(V)\) & \\
    \rowcolor{black!20}
    5
    & \(\clei \triangleq\) & \(\forall \nach \textnormal{ s.t. } \mach \approx_{\components} \nach,\)
    & \(\textnormal{ where } \components = \COMPONENTS - (\public(V) \cup \object(V))\)
    & \(\textnormal{ for all call targets } (\mach,(V,\sigma))\) \\
   \rowcolor{black!20}
    5a & & \(|\sigma| \hookrightarrow (\mach,(V,\sigma)) \simeq |\sigma| \hookrightarrow (\nach,(V,\sigma))\) & & \\
    \rowcolor{black!20}
    5b & & \(\textnormal{ and } (\mach,(V,\sigma)) ~ (|\sigma| \Uparrow \cintProp) ~ (\nach,(V,\sigma))\)
    & \(\textnormal{ where } (\mach' ~ \cintProp ~ \nach') \triangleq
    \mach' \parallel \bar{\Diamond}(\mach,\nach,\mach',\nach')\) & \\
  \end{tabular}
  \caption{Properties}
  \label{tab:props}
\end{table*}

Finally, the core property definitions are given in \cref{tab:props},
arranged to show their commonalities and distinctions. Each definition gives a criterion
quantified over states \(s\) that immediately follow call steps.
If an execution includes a transition \(s' \stepstounder{\overline{\psi}} s\)
where \(\mathbf{call} ~ \addr ~ \overline{\reg} \in \bar{\psi}\), then \(s\) is the target
of a call.
As a shorthand, we write that each property is defined
by a criterion that must hold ``for all call targets \(s\),'' or, in the case of \(\wbcf\),
``for all call steps \(s \stepstounder{} s'\).''

\paragraph*{1.~\(\wbcf\)}
Given a call step \((\mach,(V,\sigma)) \stepstounder{} (\mach',(V',\sigma'))\),
we define the predicate \(\ret\) to hold on states \(\mach''\)
whose stack pointer matches that of \(\mach\)
and whose program counter is at the next instruction. A system enjoys \(\wbcf\) if,
for every call transition, \(\ret\) holds just after the callee returns (i.e.,
the call stack shrinks).

\paragraph*{2.~\(\clri\)}
When the call target is \((\mach,(V,\sigma))\), we define the predicate \(\intProp\) to hold
on states \(\mach'\) if all elements that are both sealed in \(V\) and in the difference
set between \(\mach\) and \(\mach'\) are irrelevant. A system enjoys \(\clri\) if, for every
call, \(\intProp\) holds just after the corresponding return.

\paragraph*{3.~\(\clrc\)}
When the call target is \((\mach,(V,\sigma))\), we begin by taking an arbitrary \(\nach\)
that is a \(\components\)-variant of \(\mach\), where \(\components\) is the set of sealed elements
in \(V\). We require that two clauses hold. On line 3a, the behavior of a trace from
\((\mach,(V,\sigma))\) up to its return must match that of \((\nach,(V,\sigma))\).
On line 3b, we define a relation \(\confProp\) that relates states \(\mach'\) and \(\nach'\)
if their corrupted set (relative to \(\mach\) and \(\nach\)) is irrelevant, and require
that it hold just after the returns from the callees that start at \((\mach,(V,\sigma))\) and \((\nach,(V,\sigma))\).
A system enjoys \(\clrc\) if both clauses hold for every call.

\paragraph*{4.~\(\clec\)}
We consider the callee's private behavior to be any changes that it makes to the state
outside of legitimate channels---elements marked \(\object\) or \(\public\). The remainder
should be kept secret, which is to say, irrelevant to future execution. Similar to \(\clri\), given a call target
\((\mach,(V,\sigma))\), we define a predicate \(\cconfProp\) to hold
on states \(\mach'\) if the difference set between \(\mach\) and \(\mach'\), excluding
\(\object\) or \(\public\) locations, is irrelevant.
A system enjoys \(\clec\) if, for every call, \(\cconfProp\) holds just after the corresponding return.

\paragraph*{5.~\(\clei\)}
Callee integrity means that the caller does not influence the callee outside of legitimate
channels. The caller's influence can be seen internally, or in corrupted data on return,
just like the caller's secrets would be under \(\clrc\). So, for a call target
\((\mach,(V,\sigma))\), we take an arbitrary \(\nach\) that is a \(\components\)-variant
of \(\mach\), where \(\components\) is the set of elements that are not \(\object\)
or \(\public\). The remainder of the property is identical to \(\clrc\).

\section{Extended Code Features}
\label{sec:extensions}

The system we model in \cref{sec:example,sec:formal} is very simple, but our framework
is designed to make it easy to add support for additional code features. To support argument passing on the stack,
we just add new parameters to the existing security-relevant operations, and refine how they
update the security context. The remainder of the properties do not change at all.
To add tail-calls, we add and define a new operation, and since it is a kind of call,
we add it to the definition of call targets.
The rules for the extended security semantics are given in \cref{fig:advops}; the
rules in \cref{fig:basicops} can be recaptured by instantiating
\(\mathbf{call}\) with \(\overline{sa}\) as the empty set, and \(\mathbf{alloc}\)
with flag \(\mathbf{f}\).

\subsection{Sharing Stack Memory}
In our examples, we have presented a vision of stack safety in which
the interface between caller and callee is in the registers that pass
arguments and return values. This is frequently not the case in a realistic
setting. Arguments may be passed on the stack because there are too many
to pass in registers, as 
variadic arguments, or
because they are composite types that inherently have
pass-by-reference semantics. The caller may also pass a stack-allocated  object by reference
in the C++ style, or take its address and pass it as a pointer.

We refine our call operation to make use of the information that we have about
which stack memory locations contain arguments. The new annotation \(\overline{sa}\) is a set of
triples of a register, an offset from the value of that register, and a size.
We first define the helpful set \(\mathit{passed} ~ \overline{sa} ~ \mach\),
then extend the call operation to keep all objects in \(\mathit{passed}\) marked
as \(\object\) and seal everything else (\cref{sfig:stkargs}).

Using this mechanism, a call-by-value argument passed on the stack at an \(\SP\)-relative offset
is specified by the triple \((\SP, \mathit{off}, \mathit{sz})\).
In this case, only the immediate callee gains access to the argument location.
A C++-style call-by-reference argument where the reference is passed in \(\reg\)
is instead specified by the triple \((\reg, 0, \mathit{sz})\). Such a call-by-reference
argument could be passed through multiple calls, provided that it is in \(\overline{sa}\)
each time.

Absent the more sophisticated capability model (below), if the address of an object
is taken directly and passed as a pointer, we simply classify the object as ``public''
and give it no protection against access by other functions.
We extend the \(\mathbf{alloc}\) operation with a boolean flag, where {\bf t} indicates
that the allocation is public, and {\bf f} that it is private.
If space for multiple objects is allocated in a single step,
that step can make multiple allocation operations, each labeled appropriately.
Public objects are labeled \(\public\) rather than \(\object\), so they are
never sealed at a call (\cref{sfig:publicalloc}).
Providing more fine-grained control over sharing is desirable, but requires a considerably
more complex model. This simple model is included in our testing; we describe
an untested approach based on capabilities below.

\begin{figure*}[h]
  \begin{subfigure}{\textwidth}
    \judgmentbr[~AllocF]
        {\(\components = \mathit{range} ~ \SP ~ \mathit{off} ~ \mathit{sz} ~ \mach \cap \unsealed(V)\)}
        {\(V' = V \llbracket \addr \mapsto \object \mid \addr \in \components \rrbracket\)}
        {\(Op ~ \mach ~ (\mathbf{alloc} ~ \mathbf{f} ~ (\mathit{off, sz})) ~ (V,\sigma) = (V',\sigma)\)}
        
    \vspace{\abovedisplayskip}
        
    \judgmentbr[~AllocT]
               {\(\components = \mathit{range} ~ \SP ~ \mathit{off} ~ \mathit{sz} ~ \mach \cap
                 \unsealed(V)\)}
               {\(V' = V \llbracket \addr \mapsto \public \mid \addr \in \components \rrbracket\)}
               {\(Op ~ \mach ~ (\mathbf{alloc} ~ \mathbf{t} ~ (\mathit{off, sz})) ~ (V,\sigma) = (V',\sigma)\)}

    \vspace{\abovedisplayskip}
    
    \judgmentbr[~Dealloc]
        {\(\components = \mathit{range} ~ \SP ~ \mathit{off} ~ \mathit{sz} ~ \mach \cap (\object(V) \cup \public(V))\)}
        {\(V' = V \llbracket \addr \mapsto \unsealed | \addr \in \components \rrbracket\)}
        {\(Op ~ \mach ~ (\mathbf{dealloc} ~ (\mathit{off, sz})) ~ (V,\sigma) = (V',\sigma)\)}

    \caption{Memory Allocation}
    \label{sfig:publicalloc}
  \end{subfigure}
  \begin{subfigure}{\textwidth}

    \begin{minipage}{0.4\textwidth}
      \[\mathit{push} ~ V ~ \overline{\reg} ~ K \triangleq \lambda k .
      \begin{cases}
        \unsealed & \textnormal{if } k \in CLR \\
        \public & \textnormal{if } k \in \overline{\reg_{\mathit{args}}} \\
        \sealed & \textnormal{if } k \in \WORDS \textnormal{ and } \\
        & k \in \object(V) - \components \\
        V(k) & \textnormal{otherwise} \\
      \end{cases}      
      \]
    
      \[\mathit{passed} ~ \overline{sa} ~ \mach \triangleq \bigcup_{(\reg,\mathit{off},\mathit{sz}) \in \overline{sa}}
      \mathit{range} ~ \reg ~ \mathit{off} ~ \mathit{sz} ~ \mach\]
    \end{minipage}
    \begin{minipage}{0.6\textwidth}    
      \judgmenttwo[~Call]
        {\(\components = \mathit{passed} ~ \overline{sa} ~ \mach\)}
        {\(V' = \mathit{push} ~ V ~ \overline{\reg_{args}} ~ \components\)}
        {\(Op ~ \mach ~ (\mathbf{call} ~ \addr_{target} ~ \overline{\reg_{args}} ~ \overline{sa})
          ~ (V,\sigma) = (V',V::\sigma)\)}

    \vspace{\abovedisplayskip}
        
    \judgmenttwo[~Tailcall]
        {\(\components = \mathit{passed} ~ \overline{sa} ~ \mach\)}
        {\(V' = \mathit{push} ~ V ~ \overline{\reg_{args}} ~ \components\)}
        {\(Op ~ \mach ~ (\mathbf{tail call} ~ \addr_{target} ~ \overline{\reg_{args}} ~ \overline{sa})
          ~ (V,\sigma) = (V',\sigma)\)}
    \end{minipage}

    \caption{Calls with Argument Passing on the Stack}
    \label{sfig:stkargs}
  \end{subfigure}
  \caption{Operations supporting tail calls and argument passing on stack.}
  \label{fig:advops}
\end{figure*}

\subsection{Tail Calls}

The rule for a tail call is similar to that for a normal call.
We do not push the caller's view onto the stack,
but replace it outright. This means that a tail call does not increase the size of
the call stack, and therefore for purposes of our properties, all tail
calls will
be considered to return simultaneously when the eventual {\bf return} operation
pops the top of the stack.

Since the caller will not be returned to, it does not need integrity, but
it should still enjoy confidentiality. We set its frame to \(\unsealed\) rather
than \(\sealed\) to express this. In \cref{tab:props}, we replace
``call targets'' with ``call or tail call targets'' in \(\clrc\), \(\clec\), and \(\clei\).

\section{Provenance, Capabilities, and Protecting Objects}
\label{app:ptr}

Lastly, what if we want to express a finer-grained notion of safety, in which
stack objects are protected unless the function that owns them intentionally
passes a pointer to them? This can be thought of as a {\it capability}-based
notion of security. Capabilities are unforgeable tokens that grant access to
a region of memory, typically corresponding to valid pointers to that region.
As such, this capability safety relies on some preexisting notion of pointer
validity, i.e., {\it pointer provenance}.
Memarian et al.'s PVI \cite{provenance} (provenance via integer)
memory model is a good option: it annotates pointers with the identity of the
object they first pointed to, and propagates the annotation when the
pointer is copied and when operations are performed on it.
This constitutes a substantial addition to the security context, which is why
this enhancement is more speculative than the others, and we have not tested it.

We can model the provenance model as a trio of additional security-relevant operations: one which
declares a register to contain a valid pointer, one which transmits the provenance
of a pointer from one element to another, and one which clears the provenance
(for instance, when a pointer is modified in place in a way that makes it invalid).

In addition to the normal call stack, our security context will carry a map \(\rho\) from
elements to memory regions, represented as a base and a bound \(\context = (V, \sigma, \rho)\).
Most existing operations are extended to preserve the value of \(\rho\), while the new operations
and the call operation work as seen in \cref{fig:capops}.

\begin{figure*}
    \judgment[Promote]
             {\(\rho' = \rho[\reg_{dst} \mapsto \mathit{range} ~ \reg_{base} ~ \mathit{off} ~ \mathit{sz}]\)}
             {\(Op ~ \mach ~ (\mathbf{promote} ~ \reg_{dst} ~ (\reg_{base},\mathit{off},\mathit{sz})) ~ (V,\sigma,\rho) = (V,\sigma,\rho')\)}
             
    \vspace{\abovedisplayskip}
             
    \judgment[Clear]
             {\(\rho' = \rho[\component \mapsto \emptyset]\)}
             {\(Op ~ \mach ~ (\mathbf{clear} ~ \component) ~ (V,\sigma,\rho) = (V,\sigma,\rho')\)}
             
    \vspace{\abovedisplayskip}
             
    \judgment[Propagate]
             {\(\rho' = \rho[\component_{dst} \mapsto \rho[\component_{src}]]\)}
             {\(Op ~ \mach ~ (\mathbf{propagate} ~ \component_{src} ~ \component_{dst}) ~ (V,\sigma,\rho) = (V,\sigma,\rho')\)}
             
    \vspace{\abovedisplayskip}

    \[\mathit{reach} ~ \component ~ \rho \triangleq \{\component' | \mathit{base} \leq \component' < \mathit{bound}
    \textnormal{ where } \rho[\component] = (\mathit{base},\mathit{bound})\}\]

    \[\mathit{reach}^* ~ \components ~ \rho \triangleq \bigcup_{\component \in \components} \{ \component \} \cup \mathit{reach}^* ~ (\mathit{reach} ~ \component ~ \rho) ~ \rho\]
    \vspace{\abovedisplayskip}

    \vspace{\abovedisplayskip}
    \judgmentthree[\sc Call]
        {\(\components = \mathit{passed} ~ \overline{sa} ~ \cup ~ \overline{\reg_{args}}\)}
        {\(\components' = \mathit{reach}^* ~ \components ~ \rho\)}
        {\(V' = \mathit{push} ~ V ~ \overline{\reg_{args}} ~ K'\)}
        {\(Op ~ \mach ~ (\mathbf{call} ~ \addr_{\mathit{target}} ~ \overline{\reg_{args}} ~ \overline{sa}) ~ (V,\sigma,\rho) = (V',V::\sigma,\rho)\)}

    \vspace{\abovedisplayskip}
    \judgmentthree[\sc Tailcall]
        {\(\components = \mathit{passed} ~ \overline{sa} ~ \cup ~ \overline{\reg_{args}}\)}
        {\(\components' = \mathit{reach}^* ~ \components ~ \rho\)}
        {\(V' = \mathit{push} ~ V ~ \overline{\reg_{args}} ~ K'\)}
        {\(Op ~ \mach ~ (\mathbf{call} ~ \addr_{\mathit{target}} ~ \overline{\reg_{args}} ~ \overline{sa}) ~ (V,\sigma,\rho) = (V',\sigma,\rho)\)}

    \caption{Operations supporting provenance-based protection of passed objects}
    \label{fig:capops}
\end{figure*}

This essentially generalizes the above notion of passing: we will consider
a caller to have intentionally passed an object if that object is reachable by
a capability that has been passed to the callee. Reachability includes capabilities passed
indirectly, by being stored in an object that is in turn passed. We define
the set of reachable addresses using \(\mathit{reach*}\), the transitive closure of elements
that can be reached from the arguments of the call. The call operation in this setting
will seal only objects that are not in \(\mathit{reach*}\) nor the previously defined \(\mathit{passed}\).

In the resulting property, once an object is sealed (because its
capability has not been passed to a callee), subsequent nested calls can never unseal it.
On the other hand, an object that is passed via a pointer may be passed on indefinitely.

\section{Enforcement}
\label{sec:enforcement}

We implement and test two micro-policies inspired by
Roessler and DeHon~\cite{DBLP:conf/sp/RoesslerD18}:
{\em Depth Isolation} without lazy optimizations (DI) and with both
Lazy Tagging and Lazy Clearing optimizations (LTC).
(The connection between our properties and Roessler and DeHon's work is discussed below.)
They share a common structure: each function activation is assigned a ``color'' \(n\)
representing its identity. Stack locations belonging to that activation are
tagged \(\tagStackDepth{n}\), and while the activation is running, the tag on the
program counter (PC tag) is \(\tagPCDepth{n}\). Stack locations not part of
any activation are tagged \(\tagNoDepth\).

In DI, \(n\) always corresponds to the depth of the stack when
the function is called. A function must initialize its entire frame upon entry
in order to tag it, and then clear the frame before returning.
During normal execution, the micro-policy rules only permit load and
store operations when the target memory is tagged {\em with the same depth}
as the current {\PCname} tag, or, for store operations, if the target memory
is tagged \(\tagNoDepth\).

In LTC, a function neither initializes the frame at entry nor clears it at exist;
instead, it simply sets each location's tag to the PC tag when that location is written. It does
not check if those writes are legal! If the PC tag is \(\tagPCDepth{n}\),
then any stack location that recieves a store will be tagged \(\tagStackDepth{n}\).
On a load, the micro-policy failstops if the source memory location
is tagged \(\tagNoDepth\) or \(\tagStackDepth{n}\) for some \(n\) that
doesn't match the PC tag.

To implement this discipline, {\em blessed instruction sequences} appear at
the entry and exit of each function, which manipulate tags as just described
while performing the usual tasks of saving/restoring the return address to/from
the stack and adjusting the stack pointer. A blessed sequence uses further tags
to guarantee that the full sequence executes from the beginning---no jumping into the middle.

\paragraph*{Applicability to Roessler \& DeHon~\cite{DBLP:conf/sp/RoesslerD18}}

Roessler and DeHon (henceforward \emph{R\&D})
R\&D differentiate between memory safety policies (without lazy optimization)
and {\em data-flow integrity} policies (with lazy optimization). Our properties
are phrased in terms of data flow, and we apply them to both optimized and non-optimized
Depth Isolation.
R\&D do not attempt to define explicit formal properties, but they do list the
behaviors that they expect their data-flow integrity policies to prevent, namely:
reads from sealed objects
(our \(\clrc\)), writes to sealed objects
if they are later read (our \(\clri\)), and reads
from deallocated objects (our \(\clec\)).
They also note that Lazy Clearing prevents uninitialized reads,
which corresponds roughly to our \(\clei\).

R\&D note a flaw in Depth Isolation: because function activations
are identified by depth, a dangling pointer into a stack frame might be usable
when a new frame is allocated at the same depth. Our testing does not discover
this flaw, because we do not test address-taken objects, but it discovers a
related flaw under Lazy Tagging and Clearing that does not require
an object's address to be taken. If an activation reads a location
that was previously written by an earlier activation at the same depth, it will
violate callee confidentiality. If that location was in a caller's frame,
it also violates caller integrity and confidentiality.

They propose addressing the dangling-pointer issue by
tracking both the depth of the current activation and the static identity
of the active function. This would not eliminate all instances of this issue, but it
would require the confidentiality-violating activation to be of the same
function that wrote the data in the first place, which is a significantly higher bar.
We propose instead tracking every activation uniquely, which should eliminate the
issue entirely---and does in our tests.

\paragraph*{Protecting Registers}

R\&D do not need to protect registers, since they include the compiler in their
trusted computing base, but we target threat models that do not.
In particular, \(\clri\) requires callee-saved
registers to be saved and restored properly. We extend DI and LTC
so that callee-saved registers are also tagged with the color of the
function that is using them. In DI they are tagged as part of the entry
sequence, while in LTC they are tagged when a value is placed in them.

\section{Validation through Random Testing}
\label{sec:testing}

There are several ways to evaluate whether an enforcement mechanism enforces the above
stack safety properties. Ideally such validation would be done through formal proof over
the semantics of the enforcement-augmented machine.
However, while there are no fundamental barriers to producing such a proof,
it would be considerable work to carry out for a full ISA like RISC-V and
complex enforcement mechanisms like Roessler and DeHon's micro-policies.
We therefore choose to systematically \emph{test} their {\em Depth Isolation}
and {\em Lazy Per-Activation Tagging and Clearing} micro-policies.

We use a Coq specification of the RISC-V architecture~\cite{Bourgeat2021AMF},
extend it with a runtime monitor implementing a stack safety micro-policy,
and test it using QuickChick~\cite{Pierce:SF4}, a randomized property-based
testing framework. QuickChick works by generating
random programs, executing them, and checking that they fulfill our criteria.

Such testing is sound---it will not produce false positives---but
necessarily incomplete. We might test a flawed policy but fail to generate a
program that exploits the flaw. Additionally, detecting violations of noninterference-style
properties is dependent on choosing appropriate variant states, so it is possible
to generate a dangerous program but have it pass the test due to variant selection.
We increase our confidence in our test coverage by {\em mutation testing},
in which we intentionally inject flaws into the policies and demonstrate that testing
can find them.

\subsection{Test Generation}

To use QuickChick, we develop random test-case generators that produce
an initial RISC-V machine state tagged appropriately for the micro-policy
(see \cref{sec:enforcement}), including a code region containing a
low-level program. They also produce the meta-information about
how instructions in that program map to security-relevant operations,
which would normally be provided by the compiler.

Our generators build on the work of Hri\c{t}cu et
al. \cite{TestingNI:ICFP, DBLP:journals/jfp/HritcuLSADHPV16}, which
introduced {\em generation by execution}, a technique that produces
programs that lead to longer executions---and hopefully towards more
interesting behaviors as a result.
Each step of generation by execution takes a partially instantiated
machine state and attempts to generate an instruction
that makes sense locally (e.g., jumps go
to a potentially valid code location, loads read from a
potentially valid stack location). The generator repeats this process
for an arbitrary number of steps, or until it reaches a point where
the machine cannot step any more. Each time it generates a call or return,
it places the appropriate policy tags on the relevant instruction(s)
and records the operation.

We extend Hri\c{t}cu et al.'s technique with additional statefulness
to avoid early failstops. For example, immediately after a call, we
increase the probability of generating code that initializes any stack-allocated variables.
To allow for potential attack vectors to manifest,
the generator periodically relaxes those constraints and generates potentially ill-formed
code, such as failing to initialize variables, writing outside
of the current stack frame, or attempting an ill-formed return sequence,

\subsection{Property-based Testing}

Once a test program is generated, QuickChick tests it against a
property. A typical hyperproperty testing scheme might do this by
generating a pair of initial variant states, executing them to completion,
and comparing the results. We extend this procedure to handle the nested
nature of confidentiality.

For our setup to na\"{i}vely test the confidentiality of every call,
it would need to create a variant state at each call point, execute
it until return, then generate a post-call variant based on any tainted
values. The post-call variant would execute alongside the ``primary''
execution until the test is finished. This results in tracking a number
of variant executions that is linear in the total number of calls!

For better performance, we instead maintain a single execution that combines
all of the variants that would be spawned at returns.
So, at any given time, we need only simulate
(1) the original execution, (2) the tainted execution, and (3) one
variant execution for each call on the call stack. This approach
makes testing longer executions substantially faster, at the cost of making it
harder to identify which call is the source of a failure.

\subsection{Mutation Testing}

To ensure the effectiveness of testing against our formal properties, we
use {\em mutation testing}~\cite{JiaH11} to inject errors
(mutations) in a program that should cause the property of interest (here,
stack safety) to fail, and ensure that the testing framework can find
them. The bugs we use for our evaluation are either artificially generated
by us (deliberately weakening the micro-policy in ways that we expect
should break its guarantees), or actual bugs that we discovered through
testing our implementation. We elaborate on some such bugs below.

For example, when loading from a stack location, {\em Depth Isolation}
needs to enforce that the tag on the location being read
is $\tagStackDepth{n}$ for some number $n$ and that the tag of the
current $\PCname$ is $\tagPCDepth{n}$ for the same depth $n$. We can relax
that restriction by omitting the check (bug {\em
  LOAD\_NO\_CHECK}).
Similarly, when storing to a stack location, the correct micro-policy
needs to ensure that the tag on the memory location is either
$\tagNoDepth$ or has again the same depth as the current $\PCname$
tag. Relaxing that constraint causes violations to the integrity
property (bug {\em STORE\_NO\_CHECK}).

In additional intentional mutations, our testing catches errors in our
own implementation of the enforcement mechanism, including one interesting
bug where the initial function's frame included space allocated for its
return address, but this uninitialized (and therefore \(\tagNoDepth\)-tagged)
space was treated as private data but left unprotected. We added this to
our set of mutations as {\em HEADER\_NO\_INIT}.

For LTC, the original micro-policy, implemented as {\em PER\_DEPTH\_TAG},
fails in testing, in cases where data is leaked between sequential calls.
To round out our mutation testing we also check {\em LOAD\_NO\_CHECK},
equivalent to its counterpart in depth isolation,
and a version where stores succeed but fails to propagate the PC tag, {\em STORE\_NO\_UPDATE}.

The mean-time-to-failure (MTTF) and average number of tests for various bugs can be found in
\cref{tab:bug-table}, along with the average number of tests
it took to find the failure. Experiments were run in a desktop
machine equipped with i7-4790K CPU @ 4.0GHz with 32GB RAM.

\begin{table}[]
\centering
\begin{tabular}{c|c|c|c}
  Bug & Property Violated & Ave. MTTF (s) & Tests \\
  \hline
      {\em LOAD\_NO\_CHECK}  & Confidentiality & 24.2 & 13.3 \\
      {\em STORE\_NO\_CHECK} & Integrity & 26.9 & 26 \\
      {\em HEADER\_NO\_INIT} & Integrity & 69.5 & 76.3 \\
  \hline
  \hline
      {\em PER\_DEPTH\_TAG} & Integrity & 10.5 & 82 \\
      {\em PER\_DEPTH\_TAG} & Confidentiality & 16.85 & 88 \\
      {\em LOAD\_NO\_CHECK}  & Integrity & 8.82 & 34.3 \\
      {\em LOAD\_NO\_CHECK}  & Confidentiality & 22.55 & 127 \\
      {\em STORE\_NO\_UPDATE} & Integrity & 6.96 & 101 \\
      {\em STORE\_NO\_UPDATE} & Confidentiality & 17.34 & 11 \\
  \hline
\end{tabular}
\vspace*{1em}
\caption{MTTF for finding bugs in erroneous micro-policies: DI (top) and LTC (bottom)}
\vspace*{-2em}
\label{tab:bug-table}
\end{table}

\section{Related Work}
\label{sec:relwork}

The centrality of the function abstraction and its security are behind the
many software and hardware mechanisms proposed for its protection
\cite{Cowan+98, NagarakatteZMZ09, NagarakatteZMZ10, DeviettiBMZ08,
Kuznetsov+14, Dang+15, Shanbhogue+19, Woodruff+14, Chisnall+15,
SkorstengaardLocal, SkorstengaardSTKJFP, Georges22:TempsDesCerises,
DBLP:conf/sp/RoesslerD18, Gollapudi+23}.
%
Many enforcement techniques focus purely on \(\wbcf\);
others combine this with some degree of memory protection,
chiefly focusing on integrity.
Roessler and DeHon's {\it Depth Isolation} and {\it Lazy Tagging and Clearing}
\cite{DBLP:conf/sp/RoesslerD18} both offer protections corresponding to
\(\wbcf\), \(\clri\), and \(\clrc\), though they do not give a formal description
of this. They are generally not concerned with protecting callees.

To our knowledge, the only other line of work that aims to rigorously characterize the
security of the stack is the StkTokens-Cerise family of CHERI-enforced secure calling
conventions \cite{SkorstengaardLocal, SkorstengaardSTKJFP, Georges22:TempsDesCerises}.
The authors define stack safety as overlay semantics and related stack
safety properties, phrased in terms of logical relations instead of trace
properties.
Originally, they define an informal notion of stack safety as the combination of WBCF
and ``local state encapsulation''~\cite{SkorstengaardSTKJFP},
and describe the latter
in terms of integrity only (but it has confidentiality, equivalent
to \(\clri\) {\em and} \(\clrc\)).
StkTokens \cite{SkorstengaardSTKJFP} makes this conception of stack safety
explicit through an overlay semantics
which (1) on call mints new a stack frame from a capability representing the
available stack space, and (2) on return merges the current frame back into the
stack capability, under the assumption that there are no capabilities left on
the stack.
The underlying unary logical relation does not capture confidentiality proper,
although it does capture some of its facets.

Their latest
paper \cite{Georges22:TempsDesCerises} was inspired by the properties presented
in this paper to extend their formalism to include confidentiality through a binary logical relation. When
checking if our properties applied to their old calling convention, they noted that
it did not enforce \(\clec\), and made sure that their new version
would in addition to building it into their formalism.\footnote{A. L. Georges,
personal communication.}
%
To do so, they redesign the overlay semantics to actually pop stack frames on
return and have them disappear from the stack.
This demonstrates the benefit of our
choice to explicitly state properties in security terms: specifying security
is hard, and when the spec takes the form of a ``correct by
construction'' machine, it is easy to neglect a non-obvious security
requirement.


In terms of direct feature comparison with Georges et al.~\cite{Georges22:TempsDesCerises} (the most
recent work in the line), with the addition of confidentiality to their formalism, we
are roughly at parity in terms of the expressiveness of our properties.
We have additionally proposed callee-integrity, but it is probably the least
practical of our properties. We extend our model to tailcalls, which they do
not, and to the passing of pointers to stack objects. They discuss stack objects
and the interaction between stack and heap, but their calling convention does not
guarantee safety in the presence of pointer passing without additional checks.
We test a limited degree of pointer passing, which does not guarantee memory
safety for the passed pointer but which does not undermine the security of its
frame, and we offer an untested formalism for memory-safe passing of pointers.
On the other hand, their properties are validated by proof, while ours are
only tested.

\section{Future Work}
\label{sec:future}

We plan to test our properties against multiple enforcement mechanisms.
The top priority is capability machines, namely
CHERI \cite{DBLP:conf/sp/WatsonWNMACDDGL15}, a modern architecture designed
to provide efficient fine-grained
memory protection and compartmentalization.
We want to test the most recent work by Georges et
al.~\cite{Georges22:TempsDesCerises}, which is designed
to enforce analogues of all of our properties except for \(\clei\).

It would also be interesting to test a software enforcement approach.
Under a bounds checking discipline~
\cite{NagarakatteZMZ09}, all the pointers
in a program are extended with some disjoint metadata 
used
to gate memory accesses. These approaches enforce a form of \emph{memory safety},
and we would therefore expect them to enforce \(\clri\) and \(\clrc\). They aim
to enforce \(\wbcf\) by cutting off attacks that involve memory-safety violations,
but that may not be sufficient.
Bounds checking approaches require substantial compiler cooperation. This is not a
problem for our properties in general, but it is not very compatible with
generation-by-execution of low-level code. A better choice might be to generate
high-level code using a tool like CSmith \cite{DBLP:conf/pldi/YangCER11}, or prove the properties instead.

Several popular enforcement mechanisms are not designed to provide
absolute guarantees of security. For example, stack canaries~\cite{Cowan+98}
and shadow stacks~\cite{Dang+15,Shanbhogue+19}
are chiefly hardening techniques: they increase the difficulty
of some control-flow attacks on the stack, but cannot provide absolute
guarantees on \(\wbcf\) under a normal attacker model.
Interestingly, these are lazy enforcement mechanisms, in that
the attack may occur and be detected some time later, as long as
it is detected before it can become dangerous. That would make our
observation-based formalism a good fit for defining their security,
if we could find a formal characterization of what they do acheive
(perhaps in terms of a base machine with restricted addressing power).

We have preliminary work on extending our model to handle C++-style
exceptions, which, like tailcalls, obey only a weakened version of \(\wbcf\).
We are also exploring extensions to concurrency, starting with a model of
statically allocated co-routines.  These extensions will also require non-trivial
testing effort.  We also plan to test
the model in \cref{app:ptr} for arbitrary memory-safe pointer sharing.

\subsubsection*{Acknowledgements}

We thank the reviewers for their comments, CHR Chhak and Allison Naaktgeboren for
feedback during the writing process, and A\"ina Linn Georges for significant technical
feedback and encouragement.

This work was supported by the National Science Foundation under Grant No. 2048499, Specifying and Verifying Secure Compilation of C Code to Tagged Hardware;
by ERC Starting Grant SECOMP (715753), Efficient Formally Secure Compilers to a Tagged Architecture;
by the Deutsche Forschungsgemeinschaft (DFG, German Research Foundation) as part of the Excellence Strategy of the German Federal and State Governments, EXC 2092 CASA -- 390781972;
by NSF award \#2145649, CAREER: Fuzzing Formal Specifications,
by
NSF award \#1955610, Bringing Python Up To Speed,
and by %
NSF award \#1521523, Expeditions in Computing: The Science of Deep
  Specification.

\bibliographystyle{IEEEtran}
\bibliography{bcp.bib,local.bib}

\newcommand{\SortNoop}[1]{}
\begin{thebibliography}{10}
\providecommand{\url}[1]{#1}
\csname url@samestyle\endcsname
\providecommand{\newblock}{\relax}
\providecommand{\bibinfo}[2]{#2}
\providecommand{\BIBentrySTDinterwordspacing}{\spaceskip=0pt\relax}
\providecommand{\BIBentryALTinterwordstretchfactor}{4}
\providecommand{\BIBentryALTinterwordspacing}{\spaceskip=\fontdimen2\font plus
\BIBentryALTinterwordstretchfactor\fontdimen3\font minus
  \fontdimen4\font\relax}
\providecommand{\BIBforeignlanguage}[2]{{%
\expandafter\ifx\csname l@#1\endcsname\relax
\typeout{** WARNING: IEEEtran.bst: No hyphenation pattern has been}%
\typeout{** loaded for the language `#1'. Using the pattern for}%
\typeout{** the default language instead.}%
\else
\language=\csname l@#1\endcsname
\fi
#2}}
\providecommand{\BIBdecl}{\relax}
\BIBdecl

\bibitem{DBLP:conf/sp/RoesslerD18}
\BIBentryALTinterwordspacing
N.~Roessler and A.~DeHon, ``Protecting the stack with metadata policies and
  tagged hardware,'' in \emph{2018 {IEEE} Symposium on Security and Privacy,
  {SP} 2018, Proceedings, 21-23 May 2018, San Francisco, California,
  {USA}}.\hskip 1em plus 0.5em minus 0.4em\relax {IEEE} Computer Society, 2018,
  pp. 478--495. [Online]. Available:
  \url{https://doi.org/10.1109/SP.2018.00066}
\BIBentrySTDinterwordspacing

\bibitem{phrack96:smashingthestack}
\BIBentryALTinterwordspacing
A.~One, ``Smashing the stack for fun and profit,'' \emph{Phrack}, vol.~7,
  no.~49, November 1996. [Online]. Available:
  \url{http://www.phrack.com/issues.html?issue=49&id=14}
\BIBentrySTDinterwordspacing

\bibitem{mitre-cwe}
{MITRE Corporation}, ``Common weakness enumeration:2022 top 25 most dangerous
  software weaknesses,''
  \url{https://cwe.mitre.org/top25/archive/2022/2022_cwe_top25.html}, 2022.

\bibitem{DBLP:conf/raid/VeendCB12}
\BIBentryALTinterwordspacing
V.~van~der Veen, N.~dutt{-}Sharma, L.~Cavallaro, and H.~Bos, ``Memory errors:
  The past, the present, and the future,'' in \emph{Research in Attacks,
  Intrusions, and Defenses - 15th International Symposium, {RAID} 2012,
  Amsterdam, The Netherlands, September 12-14, 2012. Proceedings}, ser. Lecture
  Notes in Computer Science, D.~Balzarotti, S.~J. Stolfo, and M.~Cova, Eds.,
  vol. 7462.\hskip 1em plus 0.5em minus 0.4em\relax Springer, 2012, pp.
  86--106. [Online]. Available:
  \url{https://doi.org/10.1007/978-3-642-33338-5\_5}
\BIBentrySTDinterwordspacing

\bibitem{DBLP:conf/sp/SzekeresPWS13}
\BIBentryALTinterwordspacing
L.~Szekeres, M.~Payer, T.~Wei, and D.~Song, ``Sok: Eternal war in memory,'' in
  \emph{2013 {IEEE} Symposium on Security and Privacy, {SP} 2013, Berkeley, CA,
  USA, May 19-22, 2013}.\hskip 1em plus 0.5em minus 0.4em\relax {IEEE} Computer
  Society, 2013, pp. 48--62. [Online]. Available:
  \url{https://doi.org/10.1109/SP.2013.13}
\BIBentrySTDinterwordspacing

\bibitem{DBLP:conf/sp/HuSACSL16}
\BIBentryALTinterwordspacing
H.~Hu, S.~Shinde, S.~Adrian, Z.~L. Chua, P.~Saxena, and Z.~Liang,
  ``Data-oriented programming: On the expressiveness of non-control data
  attacks,'' in \emph{{IEEE} Symposium on Security and Privacy, {SP} 2016, San
  Jose, CA, USA, May 22-26, 2016}.\hskip 1em plus 0.5em minus 0.4em\relax
  {IEEE} Computer Society, 2016, pp. 969--986. [Online]. Available:
  \url{https://doi.org/10.1109/SP.2016.62}
\BIBentrySTDinterwordspacing

\bibitem{msrc-bluehat}
M.~Miller, ``Trends, challenges, and strategic shifts in the software
  vulnerability mitigation landscape,''
  \url{https://github.com/Microsoft/MSRC-Security-Research/blob/master/presentations/2019_02_BlueHatIL/},
  2019.

\bibitem{chromium-security}
{Chromium Projects}, ``Chromium security:memory safety,''
  \url{https://www.chromium.org/Home/chromium-security/memory-safety/}.

\bibitem{Cowan+98}
C.~Cowan, C.~Pu, D.~Maier, H.~Hintony, J.~Walpole, P.~Bakke, S.~Beattie,
  A.~Grier, P.~Wagle, and Q.~Zhang, ``Stackguard: Automatic adaptive detection
  and prevention of buffer-overflow attacks,'' in \emph{Proceedings of the 7th
  Conference on USENIX Security Symposium - Volume 7}, ser. SSYM’98.\hskip
  1em plus 0.5em minus 0.4em\relax USA: USENIX Association, 1998, p.~5.

\bibitem{NagarakatteZMZ09}
\BIBentryALTinterwordspacing
S.~Nagarakatte, J.~Zhao, M.~M.~K. Martin, and S.~Zdancewic, ``{SoftBound}:
  highly compatible and complete spatial memory safety for {C},'' in \emph{ACM
  SIGPLAN Conference on Programming Language Design and Implementation
  (PLDI)}.\hskip 1em plus 0.5em minus 0.4em\relax ACM, 2009, pp. 245--258.
  [Online]. Available:
  \url{http://repository.upenn.edu/cgi/viewcontent.cgi?article=1941&context=cis_reports}
\BIBentrySTDinterwordspacing

\bibitem{NagarakatteZMZ10}
\BIBentryALTinterwordspacing
------, ``{CETS}: compiler enforced temporal safety for {C},'' in \emph{9th
  International Symposium on Memory Management}.\hskip 1em plus 0.5em minus
  0.4em\relax ACM, 2010, pp. 31--40. [Online]. Available:
  \url{http://acg.cis.upenn.edu/papers/ismm10_cets.pdf}
\BIBentrySTDinterwordspacing

\bibitem{DeviettiBMZ08}
\BIBentryALTinterwordspacing
J.~Devietti, C.~Blundell, M.~M.~K. Martin, and S.~Zdancewic, ``{HardBound}:
  Architectural support for spatial safety of the {C} programming language,''
  in \emph{13th International Conference on Architectural Support for
  Programming Languages and Operating Systems}, 2008, pp. 103--114. [Online].
  Available: \url{http://acg.cis.upenn.edu/papers/asplos08_hardbound.pdf}
\BIBentrySTDinterwordspacing

\bibitem{Kuznetsov+14}
V.~Kuznetsov, L.~Szekeres, M.~Payer, G.~Candea, R.~Sekar, and D.~Song,
  ``Code-pointer integrity,'' in \emph{Proceedings of the 11th USENIX
  Conference on Operating Systems Design and Implementation}, ser.
  OSDI’14.\hskip 1em plus 0.5em minus 0.4em\relax USA: USENIX Association,
  2014, p. 147–163.

\bibitem{Dang+15}
\BIBentryALTinterwordspacing
T.~H. Dang, P.~Maniatis, and D.~Wagner, ``The performance cost of shadow stacks
  and stack canaries,'' in \emph{Proceedings of the 10th ACM Symposium on
  Information, Computer and Communications Security}, ser. ASIA CCS
  ’15.\hskip 1em plus 0.5em minus 0.4em\relax New York, NY, USA: Association
  for Computing Machinery, 2015, p. 555–566. [Online]. Available:
  \url{https://doi.org/10.1145/2714576.2714635}
\BIBentrySTDinterwordspacing

\bibitem{Shanbhogue+19}
\BIBentryALTinterwordspacing
V.~Shanbhogue, D.~Gupta, and R.~Sahita, ``Security analysis of processor
  instruction set architecture for enforcing control-flow integrity,'' in
  \emph{Proceedings of the 8th International Workshop on Hardware and
  Architectural Support for Security and Privacy}, ser. HASP ’19.\hskip 1em
  plus 0.5em minus 0.4em\relax New York, NY, USA: Association for Computing
  Machinery, 2019. [Online]. Available:
  \url{https://doi.org/10.1145/3337167.3337175}
\BIBentrySTDinterwordspacing

\bibitem{Woodruff+14}
J.~Woodruff, R.~N. Watson, D.~Chisnall, S.~W. Moore, J.~Anderson, B.~Davis,
  B.~Laurie, P.~G. Neumann, R.~Norton, and M.~Roe, ``The cheri capability
  model: Revisiting risc in an age of risk,'' in \emph{Proceeding of the 41st
  Annual International Symposium on Computer Architecuture}, ser. ISCA
  ’14.\hskip 1em plus 0.5em minus 0.4em\relax IEEE Press, 2014, p. 457–468.

\bibitem{Chisnall+15}
\BIBentryALTinterwordspacing
D.~Chisnall, C.~Rothwell, R.~N. Watson, J.~Woodruff, M.~Vadera, S.~W. Moore,
  M.~Roe, B.~Davis, and P.~G. Neumann, ``Beyond the pdp-11: Architectural
  support for a memory-safe c abstract machine,'' in \emph{Proceedings of the
  Twentieth International Conference on Architectural Support for Programming
  Languages and Operating Systems}, ser. ASPLOS ’15.\hskip 1em plus 0.5em
  minus 0.4em\relax New York, NY, USA: Association for Computing Machinery,
  2015, p. 117–130. [Online]. Available:
  \url{https://doi.org/10.1145/2694344.2694367}
\BIBentrySTDinterwordspacing

\bibitem{SkorstengaardLocal}
\BIBentryALTinterwordspacing
L.~Skorstengaard, D.~Devriese, and L.~Birkedal, ``Reasoning about a machine
  with local capabilities: Provably safe stack and return pointer management,''
  \emph{ACM Trans. Program. Lang. Syst.}, vol.~42, no.~1, Dec. 2019. [Online].
  Available: \url{https://doi.org/10.1145/3363519}
\BIBentrySTDinterwordspacing

\bibitem{SkorstengaardSTKJFP}
\BIBentryALTinterwordspacing
------, ``Stktokens: Enforcing well-bracketed control flow and stack
  encapsulation using linear capabilities,'' \emph{J. Funct. Program.},
  vol.~31, p.~e9, 2021. [Online]. Available:
  \url{https://doi.org/10.1017/S095679682100006X}
\BIBentrySTDinterwordspacing

\bibitem{Georges22:TempsDesCerises}
\BIBentryALTinterwordspacing
A.~L. Georges, A.~Trieu, and L.~Birkedal, ``Le temps des cerises: Efficient
  temporal stack safety on capability machines using directed capabilities,''
  \emph{Proc. ACM Program. Lang.}, vol.~6, no. OOPSLA1, apr 2022. [Online].
  Available: \url{https://doi.org/10.1145/3527318}
\BIBentrySTDinterwordspacing

\bibitem{Gollapudi+23}
\BIBentryALTinterwordspacing
R.~Gollapudi, G.~Yuksek, D.~Demicco, M.~Cole, G.~N. Kothari, R.~H. Kulkarni,
  X.~Zhang, K.~Ghose, A.~Prakash, and Z.~Umrigar, ``Control flow and pointer
  integrity enforcement in a secure tagged architecture,'' in \emph{2023 2023
  IEEE Symposium on Security and Privacy (SP) (SP)}.\hskip 1em plus 0.5em minus
  0.4em\relax Los Alamitos, CA, USA: IEEE Computer Society, may 2023, pp.
  1780--1795. [Online]. Available:
  \url{https://doi.ieeecomputersociety.org/10.1109/SP46215.2023.00102}
\BIBentrySTDinterwordspacing

\bibitem{pump_oakland2015}
A.~{Azevedo de Amorim}, M.~D\'en\`es, N.~Giannarakis, C.~Hri\c{t}cu, B.~C.
  Pierce, A.~Spector-Zabusky, and A.~Tolmach, ``Micro-policies: Formally
  verified, tag-based security monitors,'' in \emph{36th IEEE Symposium on
  Security and Privacy (Oakland S\&P)}.\hskip 1em plus 0.5em minus 0.4em\relax
  IEEE, May 2015.

\bibitem{sabelfeld2003language}
A.~Sabelfeld and A.~C. Myers, ``Language-based information-flow security,''
  \emph{IEEE Journal on selected areas in communications}, vol.~21, no.~1, pp.
  5--19, 2003.

\bibitem{DBLP:conf/post/AmorimHP18}
\BIBentryALTinterwordspacing
A.~{Azevedo de Amorim}, C.~Hritcu, and B.~C. Pierce, ``The meaning of memory
  safety,'' in \emph{Principles of Security and Trust - 7th International
  Conference, {POST} 2018, Held as Part of the European Joint Conferences on
  Theory and Practice of Software, {ETAPS} 2018, Thessaloniki, Greece, April
  14-20, 2018, Proceedings}, ser. Lecture Notes in Computer Science, L.~Bauer
  and R.~K{\"{u}}sters, Eds., vol. 10804.\hskip 1em plus 0.5em minus
  0.4em\relax Springer, 2018, pp. 79--105. [Online]. Available:
  \url{https://doi.org/10.1007/978-3-319-89722-6_4}
\BIBentrySTDinterwordspacing

\bibitem{Denes:VSL2014}
\BIBentryALTinterwordspacing
M.~D\'en\`es, C.~Hritcu, L.~Lampropoulos, Z.~Paraskevopoulou, and B.~C. Pierce,
  ``{QuickChick}: {P}roperty-based testing for {C}oq (abstract),'' in
  \emph{VSL}, 2014. [Online]. Available:
  \url{http://www.easychair.org/smart-program/VSL2014/index.html}
\BIBentrySTDinterwordspacing

\bibitem{Pierce:SF4}
L.~Lampropoulos and B.~C. Pierce, \emph{{QuickChick}: Property-Based Testing in
  Coq}, ser. Software Foundations series, volume 4.\hskip 1em plus 0.5em minus
  0.4em\relax Electronic textbook, Aug. 2018, version 1.0.
  \URL{http://www.cis.upenn.edu/~bcpierce/sf}.

\bibitem{pump_hasp2014}
\BIBentryALTinterwordspacing
U.~Dhawan, N.~Vasilakis, R.~Rubin, S.~Chiricescu, J.~M. Smith, T.~F. Knight,
  B.~C. Pierce, and A.~DeHon, ``{PUMP -- A Programmable Unit for Metadata
  Processing},'' in \emph{Proceedings of the 3rd International Workshop on
  Hardware and Architectural Support for Security and Privacy}, ser. HASP
  '14.\hskip 1em plus 0.5em minus 0.4em\relax New York, NY, USA: ACM, 2014.
  [Online]. Available: \url{http://www.crash-safe.org/node/32}
\BIBentrySTDinterwordspacing

\bibitem{RISC-V-CC}
R.-V. Consortium, ``Risc-v calling conventions,''
  \url{https://github.com/riscv-non-isa/riscv-elf-psabi-doc/blob/master/riscv-cc.adoc}.

\bibitem{provenance}
K.~Memarian, V.~Gomes, B.~Davis, S.~Kell, A.~Richardson, R.~Watson, and
  P.~Sewell, ``Exploring c semantics and pointer provenance,''
  \emph{Proceedings of the ACM on Programming Languages}, vol.~3, pp. 1--32, 01
  2019.

\bibitem{Bourgeat2021AMF}
T.~Bourgeat, I.~Clester, A.~Erbsen, S.~Gruetter, A.~Wright, and A.~Chlipala,
  ``A multipurpose formal risc-v specification,'' \emph{ArXiv}, vol.
  abs/2104.00762, 2021.

\bibitem{TestingNI:ICFP}
\BIBentryALTinterwordspacing
C.~Hri\c{t}cu, J.~Hughes, B.~C. Pierce, A.~Spector-Zabusky, D.~Vytiniotis,
  A.~{Azevedo de Amorim}, and L.~Lampropoulos, ``Testing noninterference,
  quickly,'' in \emph{18th ACM SIGPLAN International Conference on Functional
  Programming (ICFP)}, Sep. 2013, full version in Journal of Functional
  Programming, special issue for ICFP 2013, 26:e4 (62 pages), April 2016.
  Technical Report available as arXiv:1409.0393. [Online]. Available:
  \url{http://www.crash-safe.org/node/24}
\BIBentrySTDinterwordspacing

\bibitem{DBLP:journals/jfp/HritcuLSADHPV16}
\BIBentryALTinterwordspacing
C.~Hri\c{t}cu, L.~Lampropoulos, A.~Spector{-}Zabusky, A.~{Azevedo de Amorim},
  M.~D{\'{e}}n{\`{e}}s, J.~Hughes, B.~C. Pierce, and D.~Vytiniotis, ``Testing
  noninterference, quickly,'' \emph{J. Funct. Program.}, vol.~26, p.~e4, 2016.
  [Online]. Available: \url{https://doi.org/10.1017/S0956796816000058}
\BIBentrySTDinterwordspacing

\bibitem{JiaH11}
\BIBentryALTinterwordspacing
Y.~Jia and M.~Harman, ``An analysis and survey of the development of mutation
  testing,'' \emph{IEEE Transactions on Software Engineering}, vol.~37, no.~5,
  pp. 649--678, 2011. [Online]. Available:
  \url{http://crest.cs.ucl.ac.uk/fileadmin/crest/sebasepaper/JiaH10.pdf}
\BIBentrySTDinterwordspacing

\bibitem{DBLP:conf/sp/WatsonWNMACDDGL15}
\BIBentryALTinterwordspacing
R.~N.~M. Watson, J.~Woodruff, P.~G. Neumann, S.~W. Moore, J.~Anderson,
  D.~Chisnall, N.~H. Dave, B.~Davis, K.~Gudka, B.~Laurie, S.~J. Murdoch, R.~M.
  Norton, M.~Roe, S.~D. Son, and M.~Vadera, ``{CHERI:} {A} hybrid
  capability-system architecture for scalable software compartmentalization,''
  in \emph{2015 {IEEE} Symposium on Security and Privacy, {SP} 2015, San Jose,
  CA, USA, May 17-21, 2015}.\hskip 1em plus 0.5em minus 0.4em\relax {IEEE}
  Computer Society, 2015, pp. 20--37. [Online]. Available:
  \url{https://doi.org/10.1109/SP.2015.9}
\BIBentrySTDinterwordspacing

\bibitem{DBLP:conf/pldi/YangCER11}
\BIBentryALTinterwordspacing
X.~Yang, Y.~Chen, E.~Eide, and J.~Regehr, ``Finding and understanding bugs in
  {C} compilers,'' in \emph{Proceedings of the 32nd {ACM} {SIGPLAN} Conference
  on Programming Language Design and Implementation, {PLDI} 2011, San Jose, CA,
  USA, June 4-8, 2011}, M.~W. Hall and D.~A. Padua, Eds.\hskip 1em plus 0.5em
  minus 0.4em\relax {ACM}, 2011, pp. 283--294. [Online]. Available:
  \url{https://doi.org/10.1145/1993498.1993532}
\BIBentrySTDinterwordspacing

\end{thebibliography}

\end{document}